\documentclass[lettersize,journal]{IEEEtran}
\usepackage{amsmath,amsfonts}
\usepackage{algorithmic}
\usepackage{algorithm}
\usepackage{array}
\usepackage[caption=false,font=normalsize,labelfont=sf,textfont=sf]{subfig}
\usepackage{textcomp}
\usepackage{stfloats}
\usepackage{url}
\usepackage{verbatim}
\usepackage{graphicx}
\usepackage{cite}
\usepackage[hidelinks]{hyperref}
\begin{document}

\title{TrafficInstruct: A Multi-Task Network Traffic Dataset and Benchmark for Instruction Tuning}

\author{Zhengdao~Li, Hui~Wen, Zixuan~Wan, Weifeng~Zhang, Yan~Wu,
Benlong~Wu, Nenghai~Yu, Weiming~Zhang, and Kejiang~Chen\textsuperscript{*}%
\thanks{\hyphenpenalty=10000\exhyphenpenalty=10000 The authors are with the School of Cyberspace Security,
University of Science and Technology of China, Hefei, China
(e-mail: zhengdaoli@\allowbreak mail.\allowbreak ustc.\allowbreak edu.\allowbreak cn; harry0597@\allowbreak mail.\allowbreak ustc.\allowbreak edu.\allowbreak cn;
zixuanwan@\allowbreak mail.\allowbreak ustc.\allowbreak edu.\allowbreak cn; wf\_zhang@\allowbreak mail.\allowbreak ustc.\allowbreak edu.\allowbreak cn;
wuy4n@\allowbreak mail.\allowbreak ustc.\allowbreak edu.\allowbreak cn;
dizzylong@\allowbreak mail.\allowbreak ustc.\allowbreak edu.\allowbreak cn; ynh@\allowbreak ustc.\allowbreak edu.\allowbreak cn; zhangwm@\allowbreak ustc.\allowbreak edu.\allowbreak cn;
chenkj@\allowbreak ustc.\allowbreak edu.\allowbreak cn).\par}%
\thanks{\textsuperscript{*}Corresponding author: Kejiang Chen.}}

\markboth{TrafficInstruct}%
{A Dataset and Benchmark for Instruction-Tuned Multi-Task Network Traffic Analysis}


\maketitle

\begin{abstract}
Instruction tuning for multi-task network traffic analysis requires data resources with clearly defined collection sources, supervision targets, and evaluation conditions.
We introduce TrafficInstruct, a newly collected dataset comprising raw packet captures and 55,000 structured supervised fine-tuning records across ten tasks: malware traffic detection, botnet detection, malicious DNS-over-HTTPS detection, Web attack detection, advanced persistent threat detection, encrypted VPN detection, Tor behavior detection, encrypted application/service classification, website fingerprinting, and application classification under concept drift.
An automated workflow captures traffic from controlled software executions, attack experiments, and application activities.
Task-specific labels are supported by evidence from the corresponding collection scenarios, including process or socket attribution, activity records, and execution outcomes.
Filtering, sanitization, and task-specific transformations produce instructions, traffic observations, and structured answers.
Task-specific input preparation converts traffic observations into model prompts for baseline evaluation.
We verified the collection source of every sample and partitioned the dataset by complete collection sessions, assigning all samples from the same session to one partition.
The benchmark specifies allowed inputs, deterministic label mappings, fixed training, validation, and test partitions, and task-wise scoring rules.

We evaluate majority-class, random-forest, and ET-BERT baselines, an unadapted 27B language model, and independently trained task-specific LoRA adapters on a shared frozen base.
Each adapter is selected using its task's validation set and evaluated on the corresponding test set.
Task-specific adaptation improves classification performance, supporting the dataset's use for instruction tuning across diverse traffic-analysis tasks.
\end{abstract}

\begin{IEEEkeywords}
Network traffic analysis, multi-task benchmark, instruction tuning,
parameter-efficient fine-tuning, encrypted traffic classification.
\end{IEEEkeywords}

\section{Introduction}
\label{sec:introduction}

\IEEEPARstart{N}{etwork} traffic analysis supports security monitoring and the identification of communication behaviors.
Its tasks require different judgments, from recognizing threats to identifying applications and services in encrypted communications.
These judgments draw on related network observations, such as protocols, packet lengths, directions, and timing, but have different label semantics and observation requirements.
Instruction tuning provides a way to express such tasks through natural-language instructions and structured prediction targets~\cite{trafficllm2025}.
Using this approach across multiple tasks requires training resources that connect each instruction to an appropriate traffic representation and a clearly defined target.
The usefulness of these resources depends on how traffic is collected, how labels are assigned, and how training and evaluation samples are organized.

Existing research has advanced both traffic-model adaptation and benchmark design.
NetBench organizes seven public datasets into classification and generation tasks, with standardized preprocessing and PCAP-level partitioning before flow and packet extraction to reduce leakage between related observations~\cite{netbench2024}.
TrafficLLM combines traffic-domain tokenization, instruction and traffic tuning, and extensible parameter-efficient adaptation across diverse analysis tasks~\cite{trafficllm2025}.
These studies provide foundations for studying traffic representations and comparing models across tasks.
They also motivate complementary resources whose collection conditions, attribution evidence, and task definitions can be examined together.
Combining datasets broadens task coverage, while their original collection and labeling conditions remain relevant to interpreting results.
A common instruction format alone does not establish equivalent label evidence or evaluation conditions.
Researchers therefore need to know which observations a model may access, what evidence supports each target, and how derived samples relate to their collection sources.

We introduce TrafficInstruct, a newly collected network traffic dataset for supervised instruction tuning across ten parallel tasks: malware traffic detection, botnet detection, malicious DNS-over-HTTPS (DoH) detection, Web attack detection, advanced persistent threat (APT) detection, encrypted VPN detection, Tor behavior detection, encrypted application/service classification, website fingerprinting, and application classification under concept drift.
The complete resource contains 55,000 serialized SFT records.
Traffic is produced through controlled software executions, attack experiments, and application activities, with an automated workflow coordinating execution, capture, and quality checks.
Labels draw on evidence appropriate to each collection scenario, including process or socket attribution, activity records, and execution outcomes.
Task-specific filtering and sanitization convert the captured observations into instructions, traffic inputs, and structured targets.
Task-specific input preparation converts traffic observations into model prompts, with the experimental input conditions described in Section~\ref{subsec:experimental-setup}.
The dataset provides task-specific traffic representations, structured answers, and fixed training, validation, and test partitions.
We verified the collection source of every sample and kept all samples from the same complete collection session in one partition.

\begin{figure*}[!t]
\centering
\includegraphics[width=\textwidth]{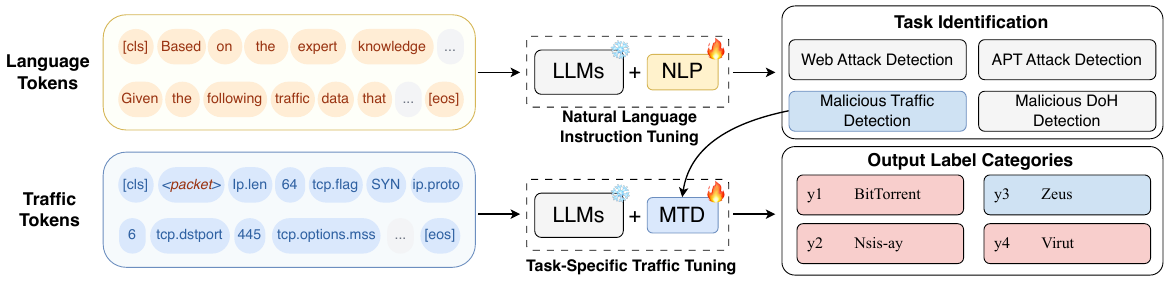}
\caption{Conceptual overview of instruction-conditioned traffic classification, inspired by the task-specific adaptation approach of TrafficLLM~\cite{trafficllm2025}. Instructions define the task, traffic observations provide its input, and the corresponding adapter specializes a frozen language-model base. Snowflake and flame symbols denote frozen and trainable components. The task panels, token notation, NLP block, and output categories are schematic; our benchmark supplies the task identity and uses independently trained task-specific adapters, with separate adapters for the plaintext and encrypted Web views.}
\label{fig:multitask-adaptation}
\end{figure*}

To support comparison across all ten tasks, our benchmark design specifies allowed inputs, deterministic label projections, fixed training/validation/test partitions, and task-wise scoring rules.
Inspired by TrafficLLM's task-specific parameter-efficient adaptation approach~\cite{trafficllm2025}, we organize the language-model baseline around a shared frozen base and independently trained task-specific PEFT modules.
Figure~\ref{fig:multitask-adaptation} presents the conceptual workflow: a natural-language instruction specifies the classification task and answer format, while the serialized traffic provides the observation to be classified.
Together, they form the input to the shared base with the adapter corresponding to the known task.
Each adapter is trained on its own task's training split and selected using that task's validation split; at inference, it generates a structured prediction in the task's label space.
The task-identification panel therefore represents supplied task context, while the token examples and output categories illustrate the information flow.
Our baseline adopts the task-specific adaptation principle rather than reproducing TrafficLLM's tokenizer or dual-stage training pipeline; the figure's NLP block is schematic and is not a separately trained module in our experiments.
Majority-class and random-forest classifiers, together with a fine-tuned ET-BERT baseline, provide complementary comparisons across the same task scope.

We evaluate classification performance across all ten tasks, compare task-specific adaptation with statistical, ET-BERT, and unadapted language-model baselines, and analyze differences between coarse and fine-grained predictions.
Together, the dataset and empirical evaluation provide a basis for studying how instruction models learn different traffic-analysis tasks from newly collected observations.

Our contributions are threefold:
\begin{itemize}
\item \textbf{A newly collected multi-task instruction resource.}
We bring together ten traffic-analysis tasks with task-specific structured targets and existing data splits, including task-specific traffic views and observation granularities.
\item \textbf{An explicit construction and evaluation methodology.}
We describe controlled collection, label evidence, filtering, and sanitization, verify sample sources, and partition samples by complete collection sessions to prevent same-session leakage across training, validation, and test sets.
\item \textbf{A task-wise empirical baseline evaluation.}
We report accuracy and Macro-F1 for statistical baselines, ET-BERT, and a 27B language model with and without task-specific LoRA, and analyze performance differences across tasks and label levels.
\end{itemize}

\section{Related Work}
\label{sec:related-work}

\subsection{Multi-Task Benchmarks and Traffic Language Models}
\label{subsec:traffic-language-models}

NetBench combines seven public datasets into fifteen classification and five generation tasks, providing a common preparation and evaluation framework~\cite{netbench2024}.
Its data preparation partitions PCAP files before extracting flows and packets, explicitly addressing correlations between observations from the same flow.
TrafficLLM combines traffic-domain tokenization, dual-stage tuning, and extensible parameter-efficient adaptation across ten analysis scenarios~\cite{trafficllm2025}.
These works establish important precedents for multi-task evaluation and traffic instruction tuning.
Our task-specific PEFT design borrows from TrafficLLM; the resource contribution concerns newly collected observations and their supervision and evaluation conditions.

Traffic pretraining provides a broader modeling context.
ET-BERT learns contextualized datagram representations from unlabeled traffic~\cite{etbert2022}, while YaTC uses masked autoencoding over a hierarchical packet/flow representation~\cite{yatc2023}.
Lens combines knowledge-guided masked-span prediction with a text-to-text formulation for classification and generation~\cite{lens2026}.
The revised netFound design incorporates protocol-aware tokenization, operational context, burst--flow attention, and restricted inputs~\cite{netfound2026}.
These approaches study reusable representations; explicit task targets and collection provenance remain relevant when preparing downstream supervision.

Evaluation methodology also matters.
Arp et al. identify pitfalls in the design and evaluation of security machine learning~\cite{arp2022}; Zhao et al. demonstrate partitioning and shortcut problems in the encrypted-traffic pipelines they examine~\cite{sugar2025}.
Such findings motivate reporting representation and partition units separately, rather than assuming all earlier resources share the same weaknesses.
In particular, PCAP files and complete activity sessions are different concepts whose correspondence must be established before comparing isolation guarantees.
Table~\ref{tab:resource-comparison} summarizes representative resources and their reported protocols without comparing scores across heterogeneous tasks.

\begin{table*}[!t]
\caption{Representative Traffic Resources and Reported Partition Units\label{tab:resource-comparison}}
\centering
\begin{tabular}{>{\raggedright\arraybackslash}p{0.11\textwidth}>{\raggedright\arraybackslash}p{0.12\textwidth}>{\raggedright\arraybackslash}p{0.15\textwidth}>{\raggedright\arraybackslash}p{0.18\textwidth}>{\raggedright\arraybackslash}p{0.10\textwidth}>{\raggedright\arraybackslash}p{0.18\textwidth}}
\noalign{\hrule height 0.8pt}
\noalign{\vskip 3pt}
Resource & Data source & Task coverage & Representation / granularity & Instruction resources & Reported partition unit \\[3pt]
\noalign{\hrule height 0.4pt}
\noalign{\vskip 3pt}
NetBench~\cite{netbench2024} & Seven public datasets & Classification and generation & Flows and packets & NR & PCAP before extraction (Sec.~3.1) \\
\noalign{\vskip 6pt}
TrafficLLM~\cite{trafficllm2025} & Public traffic datasets & Ten analysis scenarios & Traffic token sequences & Yes & 8:1:1 ratio; source groups NR in Sec.~IV-A \\
\noalign{\vskip 6pt}
CTU-13~\cite{ctu2014} & Botnet capture scenarios & Botnet / normal / background & Labeled flows; activity aggregates & NR & Scenarios in the BClus experiment (Sec.~6.2.2) \\
\noalign{\vskip 6pt}
cAPTure~\cite{capture2026} & Attack executions and replayed background & Multi-stage APT detection & Packets with attack-step labels & NR & Independent executions and background days (Sec.~4.3) \\
\noalign{\vskip 6pt}
MIRAGE-2019~\cite{mirage2019} & Human mobile-app activity & Mobile-app analysis & Packets and biflow features in capture-session JSON & NR & Collection-session evaluation grouping NR \\
\noalign{\vskip 6pt}
STAR~\cite{star2025} & HTTPS visits and browser profiles & Unseen-site retrieval & Traffic traces and semantic profiles & NR & Website identity for zero-shot evaluation (Sec.~V-A) \\
\noalign{\vskip 6pt}
UGR'16~\cite{ugr2018} & ISP traffic and injected attacks & Temporal network IDS evaluation & Anonymized NetFlow & NR & Separate calibration and test captures (Sec.~4) \\
\noalign{\vskip 6pt}
NETD / ETooL~\cite{netd2025} & Public-data recomposition; APP53 resources & Non-I.I.D. classification & Encrypted traffic representations & Yes & Controlled distribution shifts; time/version tests \\
\noalign{\vskip 6pt}
TrafficInstruct & Newly collected executions and activities & Ten classification tasks & Task-specific inputs; additional views and granularities & Yes & Complete collection sessions; sample sources individually verified \\
\noalign{\vskip 4pt}
\noalign{\hrule height 0.8pt}
\end{tabular}
\par\medskip
\parbox{\textwidth}{NR means not explicitly reported in the cited resource description or specified experimental section, rather than not performed. Instruction resources refer to an explicit task-instruction formulation. Rows describe particular published protocols, not every later use of a dataset. Our dataset uses fixed partitions based on complete collection sessions. PCAP-, scenario-, website-, and session-based partitions serve different objectives and are not ranked here.}
\end{table*}

\subsection{Threat Detection and Attribution Datasets}
\label{subsec:threat-datasets}

Threat datasets differ in both supervision semantics and the evidence connecting traffic to an activity.
USTC-TFC2016 supports malware-family and benign-application classification~\cite{ustc2017}; CTU-13 distinguishes botnet, normal, and background traffic~\cite{ctu2014}.
The latter's BClus evaluation separates capture scenarios, providing a source-isolation precedent specific to that experiment.
Threat2Traffic uses agents to reconstruct sample-dependent execution environments and collect malware traffic~\cite{threat2traffic2026}.
Grouped botnet graph metadata instead derives host/time-grouped features from CTU-13, NCC, and NCC-2~\cite{groupbot2025}.
Execution-based acquisition and graph-based derivation contribute different observations and should not be equated through record counts alone.

For encrypted DNS, CIRA-CIC-DoHBrw-2020 supports DoH detection and tunnel identification~\cite{doh2020}, while METC studies cross-network detection and tunnel-tool classification~\cite{metc2025}.
Protocol, tool, and malicious-behavior targets have different meanings even when their inputs contain similar statistics.
For Web attacks, CICIDS2017 supplies traffic from scheduled intrusion scenarios, including Web attacks~\cite{cicids2018}; the thirty-day ModSecurity resource supplies production HTTP requests selected by WAF rules~\cite{modsecurity2025}.
Rule-triggered request logs and network observations have different visibility, and an alert alone does not establish attack success.

CICAPT-IIoT connects network and provenance logs to staged attacks~\cite{cicapt2024}.
cAPTure combines attack-path executions with background traffic and supports detection-latency evaluation; it constructs training and test data from separate executions and background days~\cite{capture2026}.
These resources demonstrate the value of contextual evidence and source-aware protocols.
Our threat tracks retain task-specific attribution evidence, while the initial APT classification track uses only Flow-level benign/malicious targets; additional granularities do not imply evaluated attack-chain inference.

\subsection{Encrypted Traffic and Service Identification Datasets}
\label{subsec:encrypted-datasets}

Encrypted-traffic labels can identify communication mechanisms, application activities, or destinations.
ISCXVPN2016 and ISCXTor2016 characterize VPN and Tor traffic through time-related features and activity categories~\cite{vpn2016,tor2017}.
More recent resources investigate multiple VPN protocols~\cite{multivpn2025} and hierarchical darknet/application/behavior classification~\cite{e2emdc2025}.
Thus, behavior-level labels already have precedents; VPN or Tor membership itself is not a maliciousness judgment.

MIRAGE provides a mobile-app capture and ground-truth architecture, retaining packet and biflow information within capture-session records~\cite{mirage2019}.
Jiang et al. collect application traffic on operational network access equipment~\cite{realnet2025}.
Activity-controlled and operational-network collections offer different provenance and environmental coverage.
Our application/service track also includes service and protocol labels, so its label semantics must be distinguished from a purely mobile-app inventory.

Website fingerprinting introduces a further distinction.
AWF learns website fingerprints from Tor visit traces~\cite{awf2018}, whereas STAR aligns direct HTTPS traffic with browser-derived profiles for zero-shot website retrieval~\cite{star2025}.
STAR's website-disjoint evaluation addresses unseen identities; classification with known anonymous site labels addresses a different objective.
A complete visit trace should also be distinguished from an individual connection or a larger collection campaign.
Our benchmark therefore states task-specific observation units separately from the complete-session grouping used for its data partitions.

\subsection{Traffic Classification under Distribution Shifts}
\label{subsec:distribution-shift-datasets}

UGR'16 provides separate calibration and test captures for studying long-term evolution and periodicity in network intrusion detection~\cite{ugr2018}.
ETooL investigates instruction tuning for non-I.I.D. classification; its NETD resources include controlled recompositions of public traffic, alongside APP53 time/version tests~\cite{netd2025}.
Temporal, class-composition, and collection-session separation address different forms of generalization.

\section{Dataset Construction}
\label{sec:dataset-construction}

TrafficInstruct consists of network traffic captured during controlled software runs, attack experiments, and application activities.
Each task uses its own collection procedure, execution environment, method for identifying which activity produced the traffic, and input format.
We retain available links to captures and execution records and assign labels using the evidence available for each task.
The evaluation policy specifies which traffic fields may be provided to the model and excludes additional records used only for labeling.
The collection source of every sample is verified before assigning complete activities to data partitions.

\subsection{Controlled Execution and Traffic Collection}
\label{subsec:controlled-collection}

\begin{figure*}[!t]
\centering
\begingroup
\setlength{\unitlength}{1pt}
\begin{picture}(510,196)
\put(140,166){\framebox(230,30){\parbox{222pt}{\centering Task-specific collection\\Software runs and activities}}}
\put(10,110){\framebox(220,34){\parbox{212pt}{\centering Network observations\\PCAP and packet times}}}
\put(280,110){\framebox(220,34){\parbox{212pt}{\centering Label evidence\\Operations, sockets, execution logs}}}
\put(10,57){\framebox(220,34){\parbox{212pt}{\centering Parse, filter, and sanitize\\Flow, Flow Window, or visit Trace}}}
\put(280,57){\framebox(220,34){\parbox{212pt}{\centering Assign task-specific labels\\Check the available evidence}}}
\put(135,0){\framebox(240,34){\parbox{232pt}{\centering Structured SFT records\\Fixed data partitions}}}
\put(255,166){\line(0,-1){11}}
\put(120,155){\line(1,0){270}}
\put(120,155){\vector(0,-1){11}}
\put(390,155){\vector(0,-1){11}}
\put(120,110){\vector(0,-1){19}}
\put(390,110){\vector(0,-1){19}}
\put(120,57){\line(0,-1){12}}
\put(390,57){\line(0,-1){12}}
\put(120,45){\line(1,0){270}}
\put(255,45){\vector(0,-1){11}}
\end{picture}
\endgroup
\caption{Dataset construction: traffic supplies the observed features, while activity and execution records support labeling. The arrows show logical dependencies, rather than one execution order shared by all tasks. Samples are partitioned by complete collection sessions as described in Section~\ref{subsec:quality-audits}.}
\label{fig:construction-overview}
\end{figure*}
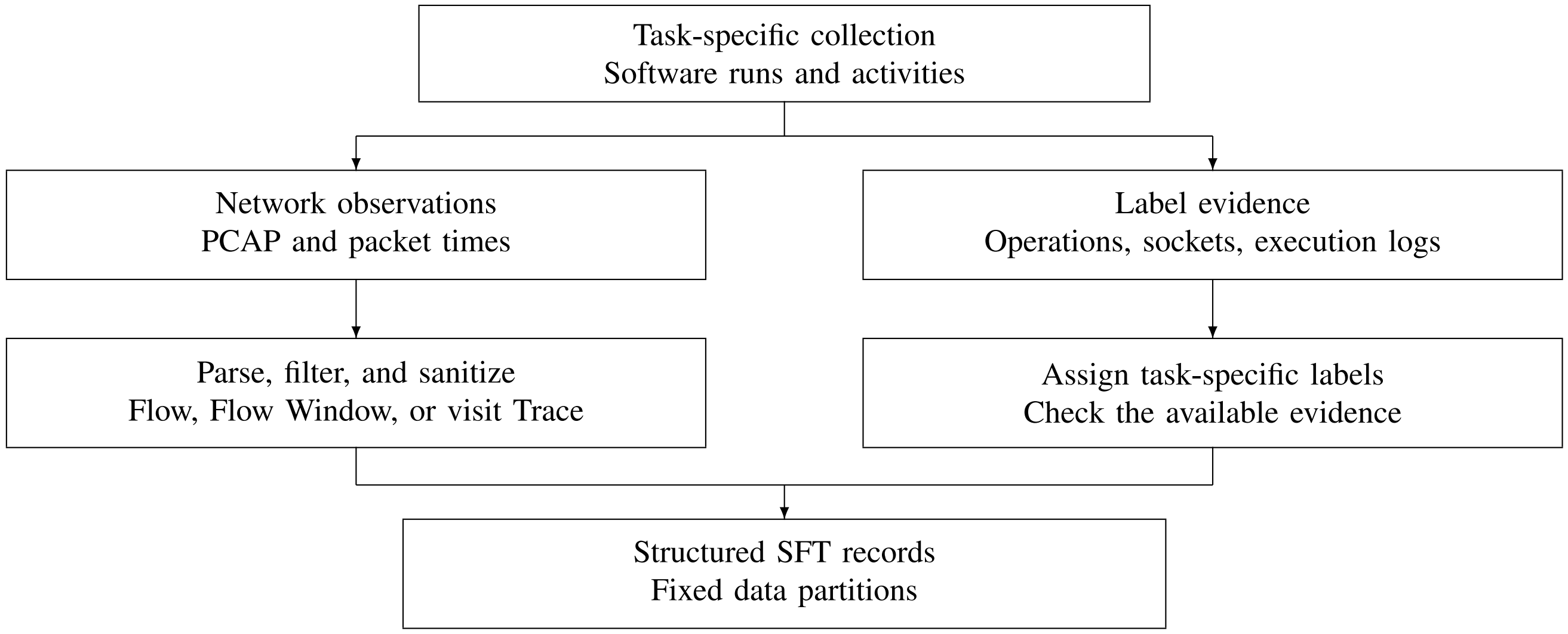

We collect traffic in isolated malware-analysis sandboxes, virtual-machine and local attack testbeds, and controlled runs of applications that access Internet services.
A \emph{collection session} is one complete capture or activity execution.
For example, one experiment may create several network connections and produce multiple Flows and SFT samples.
The collection session refers to the complete experiment or activity, rather than any one of these connections.
The collection procedure for each task is described below.
Figure~\ref{fig:construction-overview} separates network observations from records used to assign labels.
The task-specific workflows appear in Appendix~\ref{app:task-cards}.

\subsubsection{Malware Traffic Detection}
\label{subsec:collection-malware}

We obtain Windows malware samples from MalwareBazaar and execute them in a snapshot-based Windows virtual-machine testbed.
CAPE (Config And Payload Extraction), an automated malware-analysis sandbox, records process activity and network behavior during execution.
Traffic is routed through INetSim, which simulates network services within the isolated environment.
The selected runs use CAPE 2.5 and Windows 10 x64; CAPE restores the configured snapshot before each analysis and starts host-side capture before the guest analyzer executes.
Separate benign control runs execute benign network workloads under the same conditions.
Each analysis run forms a collection session with an associated packet capture and sandbox report.

We group TCP and UDP packets by transport protocol and IP-address/port endpoint pair into connection-level Sessions.
Protocol-specific timeout and TCP-instance rules distinguish repeated connections, as detailed in Appendix~\ref{app:collection-malware}.
After excluding sandbox infrastructure and network discovery traffic, we attribute Sessions using target-process connection endpoints, sandbox network observations, and application or DNS events aligned with packet times and endpoints.
Process-linked endpoints provide more specific support than endpoint-list or temporal matches.
Family labels follow the sample provider's annotations after normalization, with agreement with sandbox detections recorded when available.
Sessions lacking the required capture or attribution evidence are excluded; malicious records also require a resolved family label.
The selected data cover benign traffic and ten malware families, and records from the same source sample remain in one data split.

\subsubsection{Botnet Detection}
\label{subsec:collection-botnet}

We select Windows samples from MalwareBazaar using botnet tags or exact family signatures and analyze them with CAPE in a similarly configured, isolated Windows/INetSim testbed.
During each run, packet captures and sandbox reports record network traffic and process activity.
Separate benign control runs generate DNS and HTTP activity under the same network conditions.
We group TCP and UDP packets into bidirectional Flows and exclude sandbox control communication, inter-VM traffic, and Windows network discovery traffic.
Appendix~\ref{app:collection-botnet} gives the timeout and repeated-connection rules.

Botnet labels use target-process or descendant-process socket endpoints, process-related DNS observations, or command-and-control (C2) information extracted by CAPE.
DNS attribution uses overlap between process-related domains and sandbox DNS observations within a capture, without matching individual Flows to specific queries.
Extracted C2 information is matched by IP address without requiring a port match, or by capture-level evidence of a query for the extracted domain.
Normal denotes traffic from benign control runs.
The selected Botnet records cover five families: AsyncRAT, Lumma, SnakeKeylogger, ZLoader, and njRAT.
Records from the same source sample or control group remain in one data split.

\subsubsection{Malicious DoH Detection}
\label{subsec:collection-doh}

We collect malicious DoH traffic through controlled DNS-tunneling experiments in an isolated virtual network.
A central controller schedules three client workers to run iodine, dns2tcp, and dnscat2 with twelve scripted behaviors, covering heartbeats, jittered beacons, polling, task and result exchange, low-rate, burst, and staged uploads, configuration downloads, bidirectional task sessions, reconnection, large transfers, and mixed lifecycles.
The scripts generate synthetic messages, which the tunneling tools carry in DNS queries and responses.
A client-side \texttt{dnsproxy} forwards these queries over HTTPS to a laboratory DoH server, which terminates TLS and forwards DNS traffic to the corresponding tunnel server.
We use \texttt{tcpdump} on each client's experimental-network interface to capture bidirectional TCP/TLS traffic to the DoH server, starting before tunnel establishment and continuing through activity execution and shutdown.
Each tool--behavior execution is one collection session, with its capture, parameters, execution status, and application statistics recorded together.
These captures represent simulated malicious behaviors over a controlled DoH path.

For benign DoH, headless Chrome and Firefox visit websites drawn from the Tranco Top 10k candidate pool using AliDNS, DNSPod, 360, Switch, and CleanBrowsing.
Each visit uses a fresh browser profile and forced DoH settings, with QUIC/HTTP3 and ECH disabled.
Page loading, scrolling, and brief dwell periods generate ordinary DNS queries through the selected resolver.
We capture traffic on the browser host throughout each visit, including DoH and webpage HTTPS connections.
Separate non-DoH controls use Chrome and Firefox to visit ordinary HTTPS websites with DoH disabled.
For both browser-based collections, one website visit under a fixed browser configuration, including the selected resolver for benign DoH, and its associated PCAP define a collection session.

Collection-time checks use resolver SNI and bidirectional TLS application data to identify candidate benign DoH Flows and check for unexpected plaintext DNS or QUIC traffic.
Malicious DoH and non-DoH checks require ClientHello, ServerHello, and bidirectional TLS application data, excluding malformed Flows and those with missing TCP segments; non-DoH checks also exclude known DoH endpoints.
We retain flow-to-PCAP links and collection records, and archive failed captures separately when available.
Candidate identification and capture acceptance precede the subsequent flow filtering and SFT sample preparation.
The feature extractor operates on each PCAP separately and produces 166 statistical features per bidirectional Flow without a configured time window.
Its minimum-packet parameter is one; the TLS and bidirectional-data checks, rather than a fixed 13-packet threshold, determine which candidates qualify for SFT preparation.
Appendix~\ref{app:collection-doh} describes the collection procedure and the interpretation of fine-category labels.

\subsubsection{Web Attack Detection}
\label{subsec:collection-web}

OpenClaw is used to plan Web attack experiments against a local testbed comprising OWASP Juice Shop and controlled gateway endpoints, with communication restricted to the loopback interface.
Attack Skills and execution scripts carry out SQL injection, cross-site scripting, local and remote file inclusion, command injection, path traversal, Web scanning, and brute force.
The workflow records traffic and server-side access and audit logs, prepares an experiment report, and uses the observed results when planning subsequent experiments.
Benign activities include homepage visits, searches, resource requests, and normal failed logins; benign context requests are also interleaved with attack activities.
An HTTPS reverse proxy terminates front-end TLS before forwarding HTTP requests to a local gateway and the application.
Captures at the TLS and HTTP gateway ports provide the two views.
We group packets into bidirectional TCP Flows.
HTTP Flows are matched to request events using method/path information and bounded timing.
TLS Flows are associated offline through request anchors and connection timing within the corresponding behavior session.
A Flow containing an attack request is labeled malicious even when the attack fails; a separate connection containing only benign context remains benign.
Attack-family labels describe the attempted behavior, while response and marker checks record the local experiment outcome.
The Web attack task contains 10,000 samples: 5,000 encrypted and 5,000 plaintext records. Both views belong to the same dataset.
Appendix~\ref{app:collection-web} describes the experiment loop and sample preparation.

\subsubsection{APT Traffic Collection}
\label{subsec:collection-apt}

For APT traffic collection, we conduct 290 controlled experiments in an
isolated virtualized Windows testbed.  Before each attack experiment, network
capture is started on the monitored hosts and a CALDERA controller launches an
operation.  The controller dispatches predefined MITRE ATT\&CK abilities to
endpoint agents on the victim hosts.  The abilities cover discovery, account
and process enumeration, command and script execution, credential access,
persistence, collection and archiving, tool transfer, command and control,
exfiltration, and lateral movement.  Remote-execution and lateral-movement
variants use SMB, Windows Remote Management (WinRM), Windows Management
Instrumentation (WMI), and scheduled tasks.  The collection includes 94
successful attacks, 96 failed or blocked attempts, and 100 benign experiments
with background traffic and legitimate administrative activity.  The
orchestration environment uses MITRE CALDERA 5.3.0 with the Sandcat plugin
2.7-264-g0a35cd5 on Ubuntu 24.04.4 LTS (Linux kernel
6.8.0-136-generic).  The controller VM runs on VMware Workstation 17.5.0
(build 22583795) with virtual hardware version 21, and packet capture on the
monitored hosts uses dumpcap 4.6.8 (v4.6.8-0-ge677bf052328).

Throughout each experiment, we record network packets together with CALDERA
operation reports, ability execution times and outputs, stage timelines, host
telemetry, and experiment metadata.  After execution, the captures and records
are checked for availability, completeness, and outcome consistency.  We align
the execution timeline with the captured traffic; the stage timeline, ability
metadata, and scenario configuration provide the stage, ATT\&CK technique, and
attack-mechanism labels.  Host and network identifiers are anonymized before
serialization.

Each admitted Run starts from a restored snapshot and retains its capture, execution records, and host telemetry.
Expected attack failure is distinguished from collection failure: an interrupted operation, unusable capture, or infrastructure fault is quarantined and rerun rather than counted as a failed attack experiment.
Network events from Sysmon are matched by protocol and bidirectional endpoints; unmatched packets may inherit a stage from the bounded CALDERA execution window.
The Flow key also includes the stage identifier, so a connection spanning two stages can produce separate Flow segments.
Appendix~\ref{app:collection-apt} gives the aggregation rules and the limits of this time-based fallback.

\subsubsection{Encrypted VPN Detection}
\label{subsec:collection-vpn}

We collect VPN traffic using both an OpenVPN tunnel and a proxy service.
For OpenVPN, client version 3.9.0 (build 5008) is configured to send all host traffic through the VPN server, with UDP carrying traffic between the client and server.
For the paid proxy service, a Clash-based client uses a virtual TUN interface and global proxy routing to forward host traffic using the provider-supplied configuration; the underlying proxy protocol has not been identified.
Scripts perform Web browsing, file downloading, video streaming, and messaging; the released categories also include email and VoIP scenarios.
VPN download traffic is generated by downloading files hosted by speed-test websites.
Wireshark 4.6.2 runs on a Windows capture host and records both outgoing and incoming traffic on that host's network interface.
For Normal controls, application activities are executed directly over the ordinary network with VPN tunneling and Tor routing disabled; these activities include browsing, downloading, email, messaging, streaming, and VoIP.
The VPN and Tor tasks reuse some Normal samples within matching data splits.
Further collection details appear in Appendix~\ref{app:collection-vpn}.

\subsubsection{Tor Behavior Detection}
\label{subsec:collection-tor}

Scripts operate Tor Browser to simulate human Web browsing and to visit video webpages and play their videos.
Additional Tor Browser activities include file downloading and access to onion pages.
All Tor clients used in the collection were configured to access the Tor network through Snowflake bridges.
Each new collection session began by fully closing the previous Tor circuit and re-establishing the Tor connection through Tor Browser.
Two-host communication and file-transfer scenarios use OnionShare services.
For file transfer, one host shares files through OnionShare, and the other opens the generated onion-service address in Tor Browser and downloads the files.
For anonymous chat, one collection session spans a newly created chat room until the communication ends; Tor Browser and OnionShare are closed and restarted before the next chat session.
Wireshark 4.6.2 records both outgoing and incoming traffic at a single Windows capture host during these activities and saves it as PCAP files for later processing; in the two-host scenarios, only one host performs packet capture.
Further collection details appear in Appendix~\ref{app:collection-tor}.

\subsubsection{Encrypted Application and Service Classification}
\label{subsec:collection-app}

We execute predefined activities for 27 labels spanning applications, websites, and protocol services.
Browser automation generates website visits and resource requests, while application-specific scripts or client operations generate messaging, streaming, screen-sharing, remote-access, and file-transfer traffic.
Activity records identify the requested behavior and label, but one capture may cover only a fragment of a longer business activity.
In particular, multiple captures from a continuously running client are not assumed to be independent complete sessions.

Traffic is isolated through a dedicated proxy or loopback entry point, or sockets owned by the target process and its descendants.
The packet-capture tool \texttt{tcpdump} records traffic at the relevant loopback entry port or captures traffic for subsequent endpoint filtering.
The generic process collector follows TCP sockets, whereas the retained meeting-capture filters also include observed UDP ports.
Captures are checked for validity, label consistency, and behavior-dependent duration and size requirements; size outliers undergo manual review.
After removing proxy-establishment messages where applicable, we aggregate packets into Flows and divide them into nonoverlapping Flow Windows of at most 128 packets and 10 seconds.
Windows with fewer than three packets or traffic in only one direction are excluded.
Windows from the same source Flow or capture remain in the same split, as do captures with identical packet-structure fingerprints computed from flow membership, direction, protocol, lengths, and TCP flags.
Captures belonging to the same complete meeting or other activity are assigned to the same partition.
Appendix~\ref{app:collection-app} describes the windowing and quality rules, and Section~\ref{subsec:quality-audits} describes the collection-session partitioning rule.

\subsubsection{Website Fingerprinting}
\label{subsec:collection-wf}

We use Playwright, a browser-automation framework, to drive Chromium visits to publicly accessible HTTPS websites within a dedicated Linux network namespace.
Only one visit is active at a time, and each attempt uses a fresh browser context to avoid reusing cookies or stored state.
Before navigation, the packet-capture tool \texttt{dumpcap} starts on the client interface and records the client's TCP/UDP port 443 traffic for 30 seconds, including TLS traffic and QUIC traffic when present.
One visit capture defines a Trace and its collection session, potentially spanning multiple connections.

We accept visits that satisfy HTTP-status, destination-domain, and page-content checks and yield a valid capture containing at least 50 packets and both traffic directions.
Rejected captures and available diagnostics are retained separately to support retries or target replacement.
The selected data include 50 monitored websites represented by anonymous labels and visits to 100 other domains grouped into an unknown class.
Traces retain their source splits, and unknown domains do not overlap between splits.
Inputs combine packet events, recording inter-packet intervals and IP-packet lengths whose signs encode traffic direction, with bursts of consecutive packets traveling in the same direction.
The sequences retain at most the first 512 events and 128 bursts, while summary statistics are computed over the complete Trace for offline classification.
Real website identities and visit metadata remain separate from model inputs.
The final 5,000 records are selected deterministically within the parent split and label, preserving parent-file order; this step is not an additional random split.

\subsubsection{App Classification under Concept Drift}
\label{subsec:collection-drift}

We collect encrypted traffic from 12 Android apps, each with two predefined behaviors, A and B.
The classification label remains the app; A/B identifies the behavior performed during collection and is stored as experiment metadata.
For example, Taobao uses one versus two homepage scrolls, while Bilibili uses homepage browsing versus short video playback.
The common collection plan specifies 48 Sessions for every app, with eight independent Sessions in each of six stages: P0 uses 8A, P1--P3 use 6A+2B, 4A+4B, and 2A+6B, P4 uses 8B, and P5 repeats 8A in newly collected Sessions.
Each Session executes one registered behavior and retains the PCAP, app-operation events, and UID/socket records used to identify traffic from the target app.
The collection controller checks whether the scheduled Sessions are complete, the requested behavior executed successfully, and the captures and extracted Flows meet quality requirements.
The controller arranges retries for missing, failed, or unsuitable Sessions.
These decisions use schedule completion, execution results, and data quality; no statistical concept-drift detector is used to decide whether to collect more data.
Accepted bidirectional Flows are converted into 5,000 SFT samples with app labels and encrypted-traffic features.
The dataset assigns P0--P4 to training/validation and P5 to testing, with every collection Session assigned to one split.
Appendix~\ref{app:collection-drift} lists the registered behaviors and collection checks.

\subsection{Agent Responsibilities and Collection Workflow}
\label{subsec:agent-workflow}

The tasks use agents, experiment controllers, and execution scripts for different purposes.
In the Web task, OpenClaw supports attack planning, while attack Skills and scripts execute the planned activities and record traffic and results for subsequent planning.
In the staged mobile-app task, the controller uses predefined applications, behaviors, and environment settings to check schedule completion, behavior execution, and data quality, and arrange retries when needed.
These retries do not depend on a concept-drift detector.

Other components carry out predefined operations.
The DoH controller schedules tunnel tools and behavior scripts; CALDERA dispatches predefined abilities to endpoint agents, which execute them on the victim hosts.
Browser and application scripts perform activities for VPN, Tor, application/service classification, and website fingerprinting, while CAPE manages sandbox execution and records for malware analysis.
CALDERA endpoint agents execute predefined abilities; they are distinct from the component used for Web attack planning.
Traffic features are computed from captured packets.

\subsection{Linking Traffic to Activities and Assigning Labels}
\label{subsec:attribution-evidence}

Labels use the activity records, collection conditions, and observed results available for each task.
For malware and botnet traffic, these include process network endpoints, sandbox events, and captured network activity.
Some Botnet DNS and C2 checks apply to a whole capture: process-related domains are compared with sandbox DNS observations, and C2 IP matches do not require matching ports.
These checks do not establish a match between each Flow and a particular DNS query.
Process-linked socket endpoints provide more specific support than capture-level domain or IP matches.
Malware family labels follow normalized sample-provider annotations, with sandbox agreement recorded when available.

For malicious DoH, labels refer to controlled tool and behavior executions over the laboratory DoH path.
Browser configurations, resolver identification, and capture checks distinguish benign DoH from non-DoH controls.
For Web experiments, request roles and planned attack families determine attempt labels; observed responses and controlled markers assess experiment outcomes.
An unsuccessful attack is still an attack attempt, and an HTTP success status alone does not establish that the attack achieved its objective.
APT stage, technique, and mechanism labels also use execution timelines, ability metadata, and scenario settings.

Application/service collection isolates traffic through proxy entry points or matches sockets to the target process and its descendants.
Mobile-app collection also records operations and checks application user IDs (UIDs) and sockets.
Website-fingerprinting labels come from the controlled visit target and visit-acceptance checks; website identities remain separate from the model inputs.
VPN and Tor labels describe the communication configuration and application activity.
Their application-category labels identify the collection scenario rather than independently verified payload semantics of every Flow; Tor captures may include Snowflake support traffic.
Thus, different tasks use different records to assign labels and provide different levels of evidence.

Label names must be interpreted within each task.
Botnet Normal denotes traffic from explicitly benign control runs, while the five family labels identify attributed Botnet traffic.
VPN/Tor \texttt{Normal} denotes ordinary-network collection without VPN tunneling or Tor routing; it is not an independent assessment of maliciousness.
In the staged mobile-app task, A/B and stage are experiment metadata, while the classification label remains the app.

Reference answers are prepared separately from model inputs.
Input fields are task-specific; the APT experiments retain contextual and identifier fields in addition to Flow statistics.
Task-specific loaders prepare the classification inputs and targets used in the baseline experiments. Section~\ref{subsec:input-conversion} describes input preparation, and Section~\ref{subsec:experimental-setup} reports the actual experimental conditions.
Task-specific input preparation determines which dataset fields are supplied to the model.

\subsection{Filtering and Preparing SFT Samples}
\label{subsec:sft-representations}

Quality checks operate at different steps and on different units.
Collection checks first examine whether the intended activity ran and whether its capture is usable.
DoH checks examine TLS handshakes, bidirectional application data, capture completeness, and traffic consistent with the intended resolver or collection condition.
Website visits must pass HTTP-status, destination-domain, and page-content checks; each accepted visit capture has at least 50 packets and both traffic directions.
The mobile-app controller checks the A/B schedule, execution results, and capture and Flow quality before arranging retries or advancing the stage.

Later filtering examines the extracted samples.
Web processing removes records with missing labels, incorrect experiment matches, or invalid attack-family labels.
DoH candidates must meet the task's TLS and bidirectional-data checks; 13 packets is a reported observed minimum, not a configured admission threshold.
VPN/Tor processing removes Flows with fewer than five packets.
Application/service Flow Windows must contain at least three packets and both traffic directions.
A packet threshold for an entire webpage-visit capture is therefore different from a threshold for a Flow or a Flow Window.
Web batch checks also examine request repetition, outcome balance, and family coverage.
The fifth source batch did not pass the outcome-quota check but was retained without relabeling; these checks are distinct from packet validity and attack-attempt labels.
Appendix~\ref{app:collection-web} reports this exception rather than implying that every source batch passed every check.

The fields removed from or retained in SFT inputs also differ by task.
Malware and botnet inputs omit identifying information about their sources and raw payloads, but retain traffic statistics and selected protocol fields.
VPN, Tor, and encrypted Web inputs omit application payloads and retain packet lengths, directions, timing, and protocol fields.
Application/service inputs remove network endpoint identifiers, ports, domain names, Server Name Indication (SNI), absolute timestamps, and raw payloads.
Other tasks may retain different fields.
Raw PCAP files and execution logs are stored separately from the SFT inputs.

Each SFT record contains \texttt{instruction}, \texttt{input}, and \texttt{output}, with task-specific contents.
The records already include task instructions and supervised answers; experimental loading normally retains them, with scoped adaptations such as APT Flow binary classification described in Section~\ref{subsec:classification-targets}.
DoH inputs describe 166 statistical features as text.
Several tasks use statistics and ordered packet features from a bidirectional Flow, called a connection-level Session in the malware description.
Application/service samples cover nonoverlapping Flow Windows of at most 128 packets and 10 seconds.
Website-fingerprinting samples cover complete HTTPS-visit Traces, each potentially spanning multiple connections.
A Trace retains at most 512 packet events and 128 bursts, while summary statistics describe the complete visit.
The benchmark provides a common classification interface while retaining these task-specific observation units.
Sequence truncation limits the model's packet or event sequence and does not imply that all summary statistics use only the retained sequence.
The default evaluation is offline classification.

Dataset records and model-specific input representations serve different purposes.
APT input objects contain anonymous Flow and evidence references; additional Window inputs include a parent Run reference.
These references link records, whereas CALDERA success states describe execution outcomes.
Section~\ref{subsec:input-conversion} and Appendix~\ref{app:schema-conversion} specify which fields and target conversions the classification loaders must apply.

Web plaintext and encrypted records provide two traffic views.
The current input check finds the string \texttt{marker}, ignoring case, in 300 plaintext records.
The presence of experiment markers must be considered when interpreting classification results from plaintext inputs.
The Web data contain 5,000 HTTP/TLS Flow pairs linked through their parent run, behavior session, and request-event sets, with matched labels and original split assignments.
This event-based association is stronger than matching label order, but it is not a claim of packet-by-packet correspondence.

\subsection{APT Records at Flow, Window, and Run Levels}
\label{subsec:apt-resource}

APT data are prepared at three levels from each complete experiment, called a Run.
Packets in the PCAP files are grouped into bidirectional, stage-specific Flows, using TCP/UDP idle timeouts of 120/30 seconds.
The 60-second, nonoverlapping Windows are aligned to the Run start; a Flow crossing a boundary contributes a local packet summary to each relevant Window.
Run records summarize the complete experiment.
All records produced from the same Run are assigned to the same split.

The main benchmark asks the model to classify each APT Flow as benign or malicious.
The selected Flow records have empty packet-summary lists.
The classification input therefore contains no packet sequence and retains the available Flow statistics, contextual fields, and identifiers used in the experiments.
The converter's packet-summary capacity is not treated as evidence that packet sequences are present in these selected records.
Window and Run records are provided as additional data.
Attack-stage recognition, attack-chain reconstruction, and the benefit of combining these three levels have not been evaluated.
Experiment results, including CALDERA success states, are used for labeling or evaluation and are not given to the model as input.

\subsection{Data Quality Checks and Session-Based Splits}
\label{subsec:quality-audits}

The dataset contains 55,000 structured SFT records across ten tasks and provides fixed training, validation, and test partitions.
We verified the collection source of every sample before assigning it to a partition.

All samples from one complete capture or business activity must belong to one partition.
We call this rule \emph{collection-session-disjoint splits}.
For a sample $x$, let $g(x)$ denote its complete collection session and $s(x)$ its evaluation split:
\begin{equation}
g(x)=g(y)\ \Longrightarrow\ s(x)=s(y).
\label{eq:session-grouping}
\end{equation}
The rule applies to Flows, Flow Windows, and visit Traces according to their actual sources.
It retains additional source-sample or control-group constraints where applicable.
APT granularities inherit the parent Run assignment, and paired Web views inherit the same parent experiment assignment.
For mobile-app classification under concept drift, the dataset uses P0--P4 for training and validation and P5 for testing.
Complete Sessions are kept intact when partitioning the earlier stages into training and validation sets.
The protocol targets unseen collection sessions, without implying unseen apps, devices, attack families, or networks.

Source verification checks that samples from the same complete collection activity are kept together, including samples extracted from multiple capture fragments.
The same rule applies to linked Web views and the Flow, Window, and Run observations derived from each APT experiment.
Collection-session separation prevents leakage from the same activity across partitions; it does not imply that devices, tools, or environments are different across partitions.

We additionally check serialized inputs for exact duplicates.
For comparison, each input is converted to compact JSON with sorted object keys and hashed with SHA-256; text inputs undergo the same JSON conversion.
No split file contains repeated inputs, and no identical inputs are shared across training, validation, and test partitions.
VPN and Tor share 1,160 identical inputs in matching splits: 612 in training, 339 in validation, and 209 in testing, consistent with shared Normal traffic.
Shared controls retain matching partition assignments across the two tasks.
Input equality does not detect all similar samples or the same capture written in different formats, and separate Sessions can still share tools, environments, and activity scripts.

\section{Tasks and Evaluation}
\label{sec:tasks-and-evaluation}

This section defines the classification questions, permitted inputs, data partitions, and scoring rules for the ten-task benchmark.
This section describes the evaluation procedure used for the baseline experiments reported in Section~\ref{sec:baseline-experiments}.

\subsection{Benchmark Scope and Task Groups}
\label{subsec:benchmark-scope}

The ten tasks cover three groups: threat recognition and attribution (malware, Botnet, DoH, Web, and APT), encrypted behavior and service identification (VPN, Tor, application/service classification, and website fingerprinting), and app classification under concept drift.
Each task predicts a class from one task-specific network observation.
The common classification interface does not make their physical sample units identical: most records describe bidirectional Flows, application/service records describe nonoverlapping Flow Windows, and website-fingerprinting records describe complete visit Traces.
A collection Session is the complete capture or activity from which these observations are derived, as defined in Section~\ref{subsec:controlled-collection}.

Table~\ref{tab:task-inventory} summarizes the ten tasks, their abbreviations, observation units, and evaluated label levels.
The dataset contains 55,000 structured SFT records across ten tasks.
The Web attack task includes 5,000 plaintext and 5,000 encrypted samples.
Task-specific observation units and the inputs used for classification are described separately.

An observation may share a parent source with other records, and some Normal observations are shared by VPN and Tor.
The benchmark uses offline observations: a shortened packet sequence can coexist with statistics computed over the complete Flow or Trace.
It is therefore not an early-detection or online-monitoring benchmark.

\begin{table*}[!t]
\caption{Tasks, observation units, and classification targets. Class counts appear in parentheses. Level 1 denotes the coarse target for hierarchical tasks and the native target otherwise; a dash denotes no second semantic level.}
\label{tab:task-inventory}
\centering
\small
\setlength{\tabcolsep}{4pt}
\renewcommand{\arraystretch}{1.08}
\begin{tabular}{>{\raggedright\arraybackslash}p{0.21\textwidth}>{\centering\arraybackslash}p{0.06\textwidth}>{\raggedright\arraybackslash}p{0.09\textwidth}>{\raggedright\arraybackslash}p{0.21\textwidth}>{\raggedright\arraybackslash}p{0.15\textwidth}>{\raggedright\arraybackslash}p{0.15\textwidth}}
\hline
Task & Abbrev. & Observation & Description & Level 1 / native & Level 2 \\
\hline
Malware Traffic Detection & MTD & Connection & Detect malware traffic and identify the malware family. & Benign / malicious (2) & Benign + families (11) \\[4pt]
Botnet Detection & BND & Flow & Distinguish botnet traffic from normal traffic and identify its family. & Normal / Botnet (2) & Normal + five families (6) \\[4pt]
Malicious DoH Detection & MDD & Flow statistics & Identify malicious DNS-over-HTTPS traffic and its behavior category. & Non-DoH / benign / malicious (3) & Non-DoH + benign + behaviors (14) \\[4pt]
Web Attack Detection & WAD & Flow & Detect Web attack attempts and identify the attack category. & Benign / attack (2) & Benign + attack categories (9) \\[4pt]
APT Attack Detection & AAD & Flow segment & Classify traffic from controlled APT experiments as benign or malicious. & Benign / malicious (2) & --- \\[4pt]
Encrypted VPN Detection & EVD & Flow & Recognize VPN traffic and classify the associated network activity. & Normal / VPN (2) & Activity categories (18) \\[4pt]
Tor Behavior Detection & TBD & Flow & Recognize Tor traffic and classify the associated network activity. & Normal / Tor (2) & Activity categories (12) \\[4pt]
Encrypted App Classification & EAC & Flow Window & Identify the application, website, or protocol service generating encrypted traffic. & App/service identity (27) & --- \\[4pt]
Website Fingerprinting & WF & Visit Trace & Identify monitored websites from traffic patterns and distinguish unknown sites. & Monitored / unknown (2) & Site identities + unknown (51) \\[4pt]
Concept Drift & CD & Flow & Identify Android apps in a held-out phase of a controlled behavior schedule. & App identity (12) & --- \\
\hline
\end{tabular}
\par\smallskip
\parbox{\textwidth}{BND distinguishes Normal from Botnet at the first level and Normal from five Botnet families at the second level. WAD provides plaintext and TLS views. CD uses P0--P4 for training/validation and P5 for testing. AAD, EAC, and CD have native single-level targets.}
\end{table*}

\subsection{Labels and Classification Targets}
\label{subsec:classification-targets}

The released records contain task instructions and structured training answers.
For nine tasks, the benchmark retains these instructions and task-specific JSON output formats by default.
Each task has a designated primary label space. Scores are reported separately for each evaluated input view and label level.
A deterministic scoring function extracts the relevant class from its original answer fields.
This metric extraction leaves the training answer unchanged.
APT is the scope-specific exception: its original instruction requests stages, techniques, and evidence, so the Flow binary baseline uses a revised instruction and a derived answer containing only benign/malicious \texttt{label}.
The original APT records remain unchanged.
Appendix~\ref{app:label-projection} specifies the retained schemas, scoring fields, and targeted adaptations.

For malware, benign records map to \texttt{benign}, while malicious records map to their normalized family, producing eleven classes.
Botnet uses \texttt{fine\_label} with six classes: Normal, AsyncRAT, Lumma, SnakeKeylogger, ZLoader, and njRAT.
For first-level evaluation, Normal retains its label and the five families map to Botnet.
Web reads \texttt{category}, mapping \texttt{none} to \texttt{benign} and otherwise retaining the attack family, giving nine classes; failed attempts retain their attack class.
Its binary \texttt{label} remains in the answer and can be checked for agreement.
APT derives its binary training answer from \texttt{output.assessment.label}; its benchmark-specific instruction requests a JSON \texttt{label} with benign/malicious values.
Its stage, technique, attack outcome, evidence references, and reasoning text are not classification targets.

The DoH fine-label target contains non-DoH and benign-DoH classes plus twelve malicious behavior categories.
The coarse aliases \texttt{Non-DoH}, \texttt{Benign DoH}, and \texttt{Malicious DoH} map to the stored coarse tokens \texttt{non-doh}, \texttt{benign}, and \texttt{malicious}.
DoH retains its \texttt{label}/\texttt{category} answer; scoring maps the first two cases internally to \texttt{non-doh}/\texttt{benign} and reads the supplied category for malicious records.
A limited instruction clarification makes the stored coarse tokens and JSON field names explicit without replacing the existing classification task.
DoH evaluation uses the supplied label and category fields, reporting both three-class and fourteen-class classification results.

VPN and Tor use their released activity-category strings as the primary 18- and 12-class targets.
Fixed mappings recover VPN/Normal and Tor/Normal for supplementary coarse scores.
Application/service classification uses the 27 textual labels, and the staged mobile-app task uses the twelve package-name labels.
The original \texttt{label}/\texttt{label\_id} answers are retained; the main score reads the textual label, while integer-ID correctness and consistency are auxiliary checks.
Website fingerprinting retains \texttt{site\_label}/\texttt{is\_monitored}; its main score reads w001--w050 or \texttt{unknown} from \texttt{site\_label}.


\subsection{Collection-Session-Disjoint Splits and Stage Holdout}
\label{subsec:fixed-splits}

We verified the collection source of every sample and partitioned the dataset by complete collection sessions, following Equation~\ref{eq:session-grouping}.
All samples derived from the same complete activity are assigned exclusively to training, validation, or test, preventing same-session leakage across partitions.
The experiments use these fixed partitions.

For app classification under concept drift, the train and validation files use P0--P4 and the test file uses P5 only, with complete-session isolation.
P5 repeats behavior A in newly collected Sessions; it is neither an unseen app nor a previously unseen behavior.
The common plan specifies 48 Sessions per app, eight per stage; this planned quota is not used to infer actual per-record phase assignments.
No online drift detector, alarm-delay score, or stage-by-stage adaptation experiment is implied.

The reported results do not include Session-clustered confidence intervals.

\subsection{Allowed Inputs and Deterministic Conversion}
\label{subsec:input-conversion}

For task $t$, $G_t(x)$ denotes the complete model prompt constructed from the task instruction and permitted traffic fields of observation $x$, using the selected model's chat template.
The loader prepares the supervised answer separately, retaining the original task-specific format by default.
Only documented adaptations change the task text or answer: the APT binary projection and necessary DoH token/format clarification described above.
The same input conversion is applied to training, validation, and test observations and to adapted and unadapted language-model baselines.
Input preparation preserves the dataset records and documents the transformations applied to model prompts.
The input representations used in the experiments are identified in Section~\ref{subsec:experimental-setup}.

\subsubsection{Traffic fields}
Malware and Botnet inputs retain their supplied protocol, service-port, duration, packet/byte, directional, and packet-sequence features, including existing coarse protocol or payload-statistic buckets where available.
Encrypted Web, VPN, Tor, and mobile-app inputs retain their supplied transport/protocol fields, lengths, directions, relative times, header fields, and Flow statistics; raw payloads and endpoint identities are excluded.
Application/service inputs retain the analogous sanitized Window fields, without restoring ports, endpoints, domains, or SNI that were removed during construction.
Website inputs retain events, bursts, truncation metadata, and complete-Trace summaries, without revealing the actual website.
DoH uses the supplied 166-feature text; its measured values are not rescaled or given reconstructed formulas without checking the original extractor.
The permitted field groups are listed in Appendix~\ref{app:input-policy}.

The APT baseline uses Flow-level input objects containing network-derived statistics together with retained identifier fields, local context, and evidence references.
The selected packet-summary lists are empty, so no packet sequence is supplied.
The prediction target is benign/malicious classification for each Flow.
CALDERA outcome, stage, mechanism, technique, and confidence are excluded from input even though the released answer contains them.
This Flow-only target does not evaluate attack-chain reasoning or multi-granularity inference.
The Flow boundaries nevertheless depend on execution-stage annotations used during dataset construction.
The initial APT benchmark therefore evaluates classification of these presegmented observations from the supplied Flow-level inputs, rather than packet-only segmentation or end-to-end detection without execution metadata.

\subsubsection{Serialization and length control}
Scalar fields use a fixed key order and existing numeric precision.
Packet/event/burst sequences retain order and explicit column definitions when represented as compact arrays; schema and encoding descriptions are supplied once per prompt.
The converter records missing or inapplicable fields explicitly and does not invent zeros as replacement observations.
DoH feature names and their original text values remain associated.
Any scaling, imputation, or feature selection used by the statistical baseline is fitted on training data only.

Model inputs are tokenized using the selected model's tokenizer and processed under task-specific sequence-length limits. Section~\ref{subsec:experimental-setup} reports these limits and the experimental configuration.
If a sequence must be shortened, complete elements are retained in original order under a documented deterministic selection rule; JSON or numeric text is never cut at an arbitrary character boundary.
Summary statistics computed over the available observation before prompt-level shortening are retained and identified as such.
For VPN/Tor, this observation is limited to the processed portion of each source PCAP, as described in Appendix~\ref{app:collection-vpn}; it need not cover an entire connection's lifetime.
Truncation already present in the dataset is distinguished from additional shortening during model-input preparation.

\subsection{Task-Specific PEFT Protocol}
\label{subsec:peft-protocol}


Figure~\ref{fig:multitask-adaptation} illustrates the roles of task instructions, traffic observations, and task-specific adaptation.
The instruction specifies the classification question and expected answer format, while the serialized traffic supplies the observation to classify.
In our implementation, these components form one task-conditioned input $G_t(x)$ to a shared frozen base with the adapter for task $t$.
The two token streams in the figure distinguish information roles; they are not separate model inputs or separately trained stages.
The task is supplied by the benchmark configuration, so the ``Task Identification'' panel denotes the known task context rather than a learned routing decision.
The NLP block is schematic and does not represent an additional adapter trained in these experiments.
The displayed tokens and category names are illustrative; actual serialization, vocabularies, and structured answers follow Table~\ref{tab:task-inventory} and Section~\ref{subsec:classification-targets}.

Following the task-specific adaptation approach of TrafficLLM~\cite{trafficllm2025}, one fixed instruction-model base is shared across independently initialized task-specific adapters.
The plaintext and encrypted Web views each use a separately trained LoRA adapter.
The base parameters $\theta$ remain frozen.
Adapter $\phi_t$ is trained only on task $t$'s training partition to generate its task-specific structured answer from $G_t(x)$.
Only the APT binary baseline replaces the original rich answer with its scoped binary label.
Training loss applies to target-answer tokens, not instruction or traffic-input tokens.
Adapters are not initialized from another task's trained adapter, stacked, or selected by a learned router.
At inference, the task and input view are known, and the corresponding adapter is loaded with the shared base.


\subsection{Metrics and Scoring Rules}
\label{subsec:scoring-rules}

Each task's main score is Macro-F1 over its fixed primary label set $\mathcal{C}_t$:
\begin{equation}
\mathrm{MacroF1}_t = \frac{1}{|\mathcal{C}_t|}
\sum_{c\in\mathcal{C}_t}\frac{2\mathrm{TP}_c}{2\mathrm{TP}_c+\mathrm{FP}_c+\mathrm{FN}_c}.
\label{eq:macro-f1}
\end{equation}
A zero denominator contributes zero.
The evaluation configuration identifies its supported classes; absent support is not repaired by deleting difficult classes after inspecting model predictions.
A task's score includes every assigned test sample, not only valid model outputs or correctly identified coarse classes.

A generated answer must be one JSON object in the task's own output format, allowing surrounding whitespace only.
The parser then reads the fields required for that task's main score, rather than requiring every task to return a field named \texttt{label}.
Malformed JSON, duplicate keys, multiple answers, surrounding explanatory text, missing or wrongly typed scoring fields, and an unrecognized extracted class make the primary prediction invalid.
Invalid primary predictions add a false negative to the reference class, do not become an extra class in the macro average, and remain in the test denominator.
The scorer does not use substring rescue or repeated generation until a valid answer appears.

Primary classification scoring is distinct from full output-schema compliance.
For example, a wrong \texttt{label\_id} does not change a correctly predicted app label into a wrong app classification.
A missing auxiliary field likewise affects schema compliance rather than removing an otherwise scorable primary prediction.
The scorer does not use an auxiliary field to repair a missing or invalid primary class.
For malware's conditional main target, the required scoring fields are \texttt{maliciousness} and, for malicious predictions, \texttt{malware\_family}; for DoH they are the coarse \texttt{label} and the malicious \texttt{category} when applicable.
Coarse diagnostics derived from the primary class also retain invalid predictions as errors.

First-level classification scores derive from the primary class: malware and Web benign/malicious, Botnet Normal/Botnet, DoH non-DoH/benign/malicious, VPN/Tor their target-versus-Normal distinctions, and Website monitored/unknown.
Section~\ref{sec:baseline-experiments} reports these coarse scores alongside native single-level tasks and presents the fine-label results in a separate table for hierarchical tasks.
The fixed primary label set and validation checkpoint criterion remain unchanged.
Website fingerprinting is evaluated using Accuracy and Macro-F1 at both the monitored/unknown and fine-label levels, with invalid predictions retained as errors.


\section{Baseline Experiments}
\label{sec:baseline-experiments}

\begingroup

\subsection{Experimental Setup}
\label{subsec:experimental-setup}

We compare a training-set majority predictor, a random forest (RF), a fine-tuned ET-BERT model, an unadapted 27B language model, and the same base model with task-specific LoRA.
The evaluation addresses three questions: how well each task can be solved at its coarse or native label level, whether successful coarse classification extends to finer attribution, and how task-specific adaptation compares with statistical classification.

\paragraph{Statistical baselines}
The majority predictor chooses the mode of the training primary labels and applies the same fixed coarse mapping used for the other methods.
The RF representation is extracted from the corresponding language-model traffic input objects, excluding the answer and instruction text.


\paragraph{ET-BERT}
ET-BERT learns contextualized datagram representations through traffic pre-training~\cite{etbert2022}.
We initialize this baseline from the official pre-trained weights, with 12 Transformer layers and a hidden dimension of 768.
To accommodate the task-specific structured inputs, JSON paths and values are converted into a byte-token sequence: numerical values use training-set quantile bins, categorical values use training-set dictionaries, and long text fields contribute UTF-8 byte prefixes, encoded as hexadecimal tokens from \texttt{00} to \texttt{ff}.
This structured-byte adaptation differs from the original datagram-based input construction.
The maximum sequence length is 512 tokens for every task and Web view.
Fine-tuning uses batch size 32, 10 epochs, a learning rate of $2\times10^{-5}$, linear warm-up with ratio 0.1, and dropout 0.1.
The checkpoint with the highest validation Macro-F1 is selected, and coarse scores are obtained by mapping the fine-label predictions.

\paragraph{27B base}
The unadapted baseline uses Qwen3.8-27B directly, without task-specific fine-tuning or a LoRA adapter.
Given a task instruction and the serialized traffic observation, the model generates a structured answer from which the task's predicted class is extracted.
This baseline measures classification performance before task-specific adaptation.

\paragraph{27B + task LoRA}
This baseline adds independently trained task-specific LoRA adapters to the same Qwen3.8-27B base model. The plaintext and encrypted Web views each use a separately fine-tuned LoRA adapter and are evaluated as separate input conditions.
The base weights remain frozen, and each adapter is trained using only its corresponding task's training data.
At inference, the known task identity and input view determine which adapter is loaded.
This setting evaluates task-specific adaptation across the ten tasks.

The reference configuration specifies five epochs on eight NVIDIA RTX PRO 6000 GPUs, a per-GPU microbatch of two, and one accumulation step, giving effective batch size 16.
It uses LoRA rank 8, scaling $\alpha=32$, dropout 0.05, and \texttt{all-linear} target modules.
The reference optimizer is AdamW with learning rate $5\times10^{-5}$, a cosine schedule, warm-up ratio 0.05, and weight decay 0.01; computation is BF16, packing is disabled, and loss is applied to answer tokens.
Checkpoints are selected using validation Macro-F1 on each task’s designated selection target, without using test performance.

\paragraph{Evaluation settings}
The tasks and their label hierarchies are as described in Table~\ref{tab:task-inventory}; in addition, Table~\ref{tab:baseline-settings} lists the sequence-length limits for the 27B models, including both Web views. ET-BERT uses its separate 512-token limit.
Our experimental conditions contain 3000/1000/1000 training/validation/test observations.
WAD is evaluated separately on compressed plaintext HTTP (WAD-Plain) and encrypted TLS traffic (WAD-Enc), while AAD retains Flow features, context, and identifiers. CD uses P0–P4 for training/validation and P5 for testing, with application identity as its classification target. These input and split conditions define the scope of the reported results.

Acc. and Macro-F1 are reported as percentages on each task's test set.
For the generative 27B models, under the strict JSON scoring protocol in Section~\ref{sec:tasks-and-evaluation}, invalid primary predictions count as errors, while full-schema and auxiliary-field correctness are assessed separately.
For hierarchical tasks, the paired evaluation uses the same selected checkpoint and test predictions at both label levels.


\begin{table}[!t]
\caption{Task-specific maximum sequence lengths for the 27B models. ET-BERT uses 512 tokens for all tasks and both Web views. Task abbreviations follow Table~\ref{tab:task-inventory}.}
\label{tab:baseline-settings}
\centering
\small
\setlength{\tabcolsep}{3pt}
\renewcommand{\arraystretch}{1.18}
\begin{tabular}{@{}>{\centering\arraybackslash}m{0.24\columnwidth}>{\centering\arraybackslash}m{0.48\columnwidth}@{}}
\hline
\parbox[c][32pt][c]{\linewidth}{\centering\bfseries Task} & \parbox[c][32pt][c]{\linewidth}{\centering\bfseries Maximum sequence\\length (tokens)} \\
\hline
MTD & 4096 \\
BND & 4096 \\
MDD & 4096 \\
WAD-Plain & 8192 \\
WAD-Enc & 8192 \\
AAD & 1024 \\
EVD & 8192 \\
TBD & 6144 \\
EAC & 8192 \\
WF & 8192 \\
CD & 8192 \\
\hline
\end{tabular}
\end{table}

\subsection{First-Level and Single-Level Classification}
\label{subsec:baseline-level1}

Table~\ref{tab:baseline-level1-results} reports accuracy (Acc.) and Macro-F1 for all ten tasks, with separate rows for the two Web views.
For hierarchical tasks, the first level is benign/malicious for MTD and WAD, Normal/Botnet for BND, non-DoH/benign/malicious for MDD, Normal/VPN for EVD, Normal/Tor for TBD, and monitored/unknown for WF.
AAD retains its Flow-level benign/malicious target, while EAC and CD retain their native app/service and app-identity labels.
Numeric app IDs encode those identities and do not define an additional semantic level.
The evaluation protocol derives coarse predictions from the extracted primary class; an invalid primary output remains invalid after projection and cannot be repaired using an auxiliary field.

\paragraph{Task-specific adaptation improves classification performance}
LoRA achieves higher Macro-F1 than the base model on all ten tasks in Table~\ref{tab:baseline-level1-results}.
For example, MTD increases from 53.8\% to 93.5\%, BND from 54.3\% to 98.1\%, and EAC from 1.7\% to 70.4\%.
CD increases from 7.0\% to 87.3\% on P5.
These comparisons indicate the benefit of task-specific supervision.

\paragraph{The benefit over RF varies substantially by task}
LoRA achieves the highest Macro-F1 in ten of the eleven first-level or native-label rows; ET-BERT leads on MDD.
Based on the displayed table values, the margin over RF ranges from 0.6 percentage points on EVD to 12.8 points on EAC.
For EVD, the scores are 85.6\% and 85.0\%, respectively, showing that a statistical baseline is already competitive at the coarse level.
For EAC, LoRA reaches 70.4\% versus RF's 57.6\%; for CD, it reaches 87.3\% versus 76.2\%.
These task-dependent margins support reporting individual tasks rather than summarizing heterogeneous label spaces and input conditions with one overall mean.
No significance claim is made from these single-run differences.

\paragraph{Comparison with ET-BERT}
ET-BERT reaches 97.5\% Macro-F1 on MDD, compared with LoRA's 97.4\%.
On CD, ET-BERT has slightly higher accuracy (86.9\% versus 86.8\%), while LoRA has higher Macro-F1 (87.3\% versus 86.5\%).
LoRA has higher coarse/native-label Macro-F1 on the remaining tasks and both Web views, including 94.4\% versus 86.8\% on encrypted Web traffic.

\paragraph{Accuracy can hide weak class-balanced performance}
The AAD majority predictor obtains 89.5\% accuracy but only 47.2\% Macro-F1, illustrating the effect of its imbalanced binary labels.
The discrepancy is particularly pronounced for WF: both the majority predictor and RF reach 97.5\% accuracy but 49.4\% Macro-F1 at the monitored/unknown level.
LoRA increases WF accuracy only to 97.7\%, while Macro-F1 reaches 56.8\%.
Thus, a high coarse accuracy does not establish reliable recognition of unknown websites, and class-specific support and errors remain necessary for interpreting this task.

\begin{table*}[!t]
\caption{First-level and single-level test classification under the conditions described in Section~\ref{subsec:experimental-setup}. Each method reports Acc. and Macro-F1 as percentages. Hierarchical tasks use the fixed coarse projection; tasks without a hierarchy retain their native labels. Bold values indicate the best result for each task and metric, including ties.}
\label{tab:baseline-level1-results}
\centering
\small
\setlength{\tabcolsep}{3pt}
\renewcommand{\arraystretch}{1.18}
\begin{tabular*}{0.9\textwidth}{@{}>{\raggedright\arraybackslash}p{0.082\textwidth}@{\hspace{8pt}\extracolsep{\fill}}>{\centering\arraybackslash}p{0.071\textwidth}@{\hspace{2pt}\extracolsep{0pt}}>{\centering\arraybackslash}p{0.071\textwidth}@{\hspace{8pt}\extracolsep{\fill}}>{\centering\arraybackslash}p{0.071\textwidth}@{\hspace{2pt}\extracolsep{0pt}}>{\centering\arraybackslash}p{0.071\textwidth}@{\hspace{8pt}\extracolsep{\fill}}>{\centering\arraybackslash}p{0.071\textwidth}@{\hspace{2pt}\extracolsep{0pt}}>{\centering\arraybackslash}p{0.071\textwidth}@{\hspace{8pt}\extracolsep{\fill}}>{\centering\arraybackslash}p{0.071\textwidth}@{\hspace{2pt}\extracolsep{0pt}}>{\centering\arraybackslash}p{0.071\textwidth}@{\hspace{8pt}\extracolsep{\fill}}>{\centering\arraybackslash}p{0.071\textwidth}@{\hspace{2pt}\extracolsep{0pt}}>{\centering\arraybackslash}p{0.071\textwidth}@{}}
\hline
\noalign{\vskip 3pt}
 & \multicolumn{2}{c}{\textbf{Majority class}} & \multicolumn{2}{c}{\textbf{Random forest}} & \multicolumn{2}{c}{\textbf{ET-BERT}} & \multicolumn{2}{c}{\shortstack{\textbf{Qwen3.8 27B}\\\textbf{base}}} & \multicolumn{2}{c}{\shortstack{\textbf{Qwen3.8 27B}\\\textbf{+ task LoRA}}} \\
\textbf{Task} & Acc. & Macro-F1 & Acc. & Macro-F1 & Acc. & Macro-F1 & Acc. & Macro-F1 & Acc. & Macro-F1 \\
\hline
MTD & 50.0 & 33.3 & 89.2 & 89.1 & 92.6 & 92.6 & 60.9 & 53.8 & \textbf{93.5} & \textbf{93.5} \\
BND & 50.0 & 33.3 & 91.3 & 88.1 & 96.5 & 96.5 & 61.0 & 54.3 & \textbf{98.8} & \textbf{98.1} \\
MDD & 33.3 & 16.7 & 92.1 & 90.8 & 97.5 & \textbf{97.5} & 53.2 & 47.7 & \textbf{97.8} & 97.4 \\
WAD-Plain & 50.5 & 33.6 & 89.2 & 89.2 & 92.2 & 92.2 & 87.8 & 87.7 & \textbf{100.0} & \textbf{100.0} \\
WAD-Enc & 50.5 & 33.6 & 86.4 & 85.9 & 86.8 & 86.8 & 50.5 & 33.6 & \textbf{94.4} & \textbf{94.4} \\
AAD & 89.5 & 47.2 & 94.3 & 82.2 & 88.9 & 69.5 & 91.8 & 71.8 & \textbf{98.1} & \textbf{94.4} \\
EVD & 50.0 & 33.3 & 85.2 & 85.0 & 82.5 & 82.5 & 45.6 & 38.0 & \textbf{85.6} & \textbf{85.6} \\
TBD & 50.0 & 33.3 & 91.8 & 91.2 & 92.8 & 91.3 & 50.0 & 33.3 & \textbf{96.0} & \textbf{96.0} \\
EAC & 3.7 & 0.3 & 57.3 & 57.6 & 49.1 & 49.1 & 6.8 & 1.7 & \textbf{71.0} & \textbf{70.4} \\
WF & 97.5 & 49.4 & 97.5 & 49.4 & 97.6 & 53.2 & 90.2 & 46.4 & \textbf{97.7} & \textbf{56.8} \\
CD & 10.4 & 1.6 & 80.2 & 76.2 & \textbf{86.9} & 86.5 & 10.8 & 7.0 & 86.8 & \textbf{87.3} \\
\hline
\end{tabular*}
\par\smallskip
\parbox{0.9\textwidth}{\footnotesize WAD-Plain and WAD-Enc denote the plaintext HTTP and encrypted TLS views of the same Web task; CD uses the P5 test phase. ET-BERT uses structured-byte inputs. Hierarchical and single-level tasks have different label meanings, so this table does not impose one shared binary target. BND evaluates Normal/Botnet classification at the first level and Normal plus five families at the second level.}
\end{table*}

\subsection{Second-Level Classification}
\label{subsec:baseline-level2}

Table~\ref{tab:baseline-level2-results} reports finer classification for the seven hierarchical tasks, with the two Web views evaluated separately.
MTD distinguishes benign traffic and malware families; BND retains Normal alongside five families; MDD retains non-DoH and benign traffic alongside malicious behavior categories.
WAD uses attack families or benign, EVD/TBD use activity categories, and WF uses site identity or unknown.
AAD, EAC, and CD have no additional semantic level and are omitted from this table.
The paired evaluation protocol uses the same checkpoint, test observations, and predictions at both levels, without independent model selection for the second table.
Fine-label evaluation includes the full test population, including terminal non-threat and unknown classes, rather than conditioning on correct coarse predictions.

\paragraph{Coarse success does not imply fine attribution}
LoRA's Macro-F1 is 93.5\% at the MTD detection level but 52.7\% at the family level.
MDD similarly has 97.4\% coarse Macro-F1 and 52.5\% fine-label Macro-F1; TBD has 96.0\% and 52.8\%, respectively.
These results identify fine attribution as an additional challenge that would be obscured by reporting only broad detection or traffic-type recognition.
BND retains a higher fine-label Macro-F1 of 86.2\%, and plaintext WAD reaches 99.3\% compared with its 100.0\% coarse score.
Because the two levels average over different label sets, their Macro-F1 difference is not a conditional error rate or an exact decomposition of the coarse errors.

\paragraph{The ranking changes at the fine level}
At the fine-label level, LoRA achieves higher Macro-F1 than RF on BND, MDD, and both Web views, while RF performs better on the other four tasks.
Across all five baselines, ET-BERT achieves the highest MTD fine-label accuracy and Macro-F1 (78.0\% and 62.3\%); LoRA leads in fine-label Macro-F1 on BND, MDD, and both Web views, while RF leads on EVD, TBD, and WF.
For example, RF obtains 65.7\% on TBD and 65.2\% on WF, compared with LoRA's 52.8\% and 53.5\%.
The corresponding gaps of 12.9 and 11.7 percentage points show that successful task adaptation does not uniformly outperform statistical classification for activity or site identification.
The experiment compares the reported representations and training histories, so it does not identify which individual feature or optimization choice causes these differences.

\paragraph{Plaintext and encrypted Web views}
LoRA obtains 94.4\% coarse Macro-F1 and 80.3\% fine-label Macro-F1 on WAD-Enc, compared with 100.0\% and 99.3\% on WAD-Plain.
On WAD-Enc, its fine-label Macro-F1 exceeds RF's 62.7\% and ET-BERT's 62.1\%.
The lower encrypted-view scores highlight the difference between the two input representations without attributing the entire gap to encryption alone.

\paragraph{Fine-label accuracy and Macro-F1 provide complementary evidence}
On MTD, LoRA has higher accuracy than RF (74.0\% versus 70.9\%) but lower Macro-F1 (52.7\% versus 59.6\%).
On EVD, both methods obtain 59.0\% accuracy, yet RF has the higher Macro-F1 (54.4\% versus 53.1\%).
These examples show why the benchmark retains a class-balanced metric alongside overall accuracy.
WF also illustrates why the two levels should be interpreted separately: RF's site-level Macro-F1 is 65.2\%, while its monitored/unknown Macro-F1 is 49.4\%.
Unknown is one class among many at the fine level but one of only two at the coarse level; the two macro averages therefore assign different relative weight to unknown-site errors.

\begin{table*}[!t]
\caption{Second-level test classification for tasks with a label hierarchy. Acc. and Macro-F1 are percentages over each task's full test set and fixed fine-label vocabulary. Single-level tasks are omitted by design. Bold values indicate the best result for each task and metric, including ties.}
\label{tab:baseline-level2-results}
\centering
\small
\setlength{\tabcolsep}{3pt}
\renewcommand{\arraystretch}{1.18}
\begin{tabular*}{0.9\textwidth}{@{}>{\raggedright\arraybackslash}p{0.082\textwidth}@{\hspace{8pt}\extracolsep{\fill}}>{\centering\arraybackslash}p{0.071\textwidth}@{\hspace{2pt}\extracolsep{0pt}}>{\centering\arraybackslash}p{0.071\textwidth}@{\hspace{8pt}\extracolsep{\fill}}>{\centering\arraybackslash}p{0.071\textwidth}@{\hspace{2pt}\extracolsep{0pt}}>{\centering\arraybackslash}p{0.071\textwidth}@{\hspace{8pt}\extracolsep{\fill}}>{\centering\arraybackslash}p{0.071\textwidth}@{\hspace{2pt}\extracolsep{0pt}}>{\centering\arraybackslash}p{0.071\textwidth}@{\hspace{8pt}\extracolsep{\fill}}>{\centering\arraybackslash}p{0.071\textwidth}@{\hspace{2pt}\extracolsep{0pt}}>{\centering\arraybackslash}p{0.071\textwidth}@{\hspace{8pt}\extracolsep{\fill}}>{\centering\arraybackslash}p{0.071\textwidth}@{\hspace{2pt}\extracolsep{0pt}}>{\centering\arraybackslash}p{0.071\textwidth}@{}}
\hline
\noalign{\vskip 3pt}
 & \multicolumn{2}{c}{\textbf{Majority class}} & \multicolumn{2}{c}{\textbf{Random forest}} & \multicolumn{2}{c}{\textbf{ET-BERT}} & \multicolumn{2}{c}{\shortstack{\textbf{Qwen3.8 27B}\\\textbf{base}}} & \multicolumn{2}{c}{\shortstack{\textbf{Qwen3.8 27B}\\\textbf{+ task LoRA}}} \\
\textbf{Task} & Acc. & Macro-F1 & Acc. & Macro-F1 & Acc. & Macro-F1 & Acc. & Macro-F1 & Acc. & Macro-F1 \\
\hline
MTD & 50.0 & 6.1 & 70.9 & 59.6 & \textbf{78.0} & \textbf{62.3} & 51.3 & 8.4 & 74.0 & 52.7 \\
BND & 50.0 & 11.1 & 87.5 & 81.0 & 94.0 & 85.1 & 56.6 & 18.7 & \textbf{96.5} & \textbf{86.2} \\
MDD & 33.3 & 3.6 & 73.3 & 46.5 & \textbf{78.9} & 43.4 & 33.3 & 3.8 & \textbf{78.9} & \textbf{52.5} \\
WAD-Plain & 50.5 & 7.5 & 75.8 & 59.7 & 78.8 & 61.0 & 86.9 & 88.3 & \textbf{99.7} & \textbf{99.3} \\
WAD-Enc & 50.5 & 7.5 & 72.4 & 62.7 & 76.1 & 62.1 & 50.5 & 7.5 & \textbf{88.6} & \textbf{80.3} \\
EVD & 8.9 & 0.9 & \textbf{59.0} & \textbf{54.4} & 56.0 & 51.5 & 18.0 & 8.8 & \textbf{59.0} & 53.1 \\
TBD & 9.8 & 1.5 & \textbf{65.9} & \textbf{65.7} & 43.7 & 43.1 & 17.8 & 7.0 & 52.6 & 52.8 \\
WF & 1.9 & 0.1 & \textbf{65.7} & \textbf{65.2} & 54.4 & 53.9 & 7.2 & 6.4 & 57.2 & 53.5 \\
\hline
\end{tabular*}
\par\smallskip
\end{table*}

\subsection{Benchmark Observations and Reporting Limits}
\label{subsec:benchmark-observations}

Taken together, the results motivate a benchmark that preserves both coarse decisions and fine attribution.
Task-specific LoRA achieves higher Macro-F1 than the base model across all ten tasks and both Web views, and leads ten of the eleven first-level or native-label rows.
ET-BERT leads MDD coarse Macro-F1 and MTD fine-label Macro-F1, adding a complementary pre-trained traffic-model comparison.
At the fine-label level, LoRA's advantage over RF varies by task.
The resource therefore supports more than a single detection score: it exposes distinctions among threat detection, family or behavior attribution, activity recognition, and site identification.
These observations characterize the benchmark results and do not establish a new adaptation algorithm or universal language-model superiority.

CD achieves 87.3\% Macro-F1 on the P5 holdout, compared with 7.0\% for the base model and 76.2\% for RF.
It does not measure the performance of a drift detector, and a validation--test difference alone would not isolate the effect of drift from changes in class composition or other phase-specific conditions.
WAD's near-perfect classification applies to the compressed plaintext HTTP view; encrypted Web results are evaluated separately.
AAD retains contextual and identifier fields, so potential shortcut features have not been ruled out; their presence alone is not proof of cross-session leakage.
BND evaluates Normal traffic and five Botnet families under the fixed task partitions.
MDD fine-label scores measure prediction of the supplied category names; their correspondence to the recorded collection behaviors has not been independently verified.

Under the scoring protocol, invalid primary outputs remain errors, while auxiliary fields do not alter an otherwise valid primary classification.
The experiments report Accuracy and Macro-F1 at the applicable label levels.
The limited unknown-site support does not substantiate claims about very-low-false-positive-rate deployment.
A single seed does not establish statistical significance, and complete collection-session identifiers are needed for session-grouped confidence intervals.
The results should consequently be interpreted within the stated task definitions, input conditions, and verification limits.

\endgroup

\section{Limitations and Access}
\label{sec:limitations-and-access}
\mbox{}\par

\subsection{Limitations}
\label{subsec:environment-limitations}

Our dataset is collected through controlled software executions, attack experiments, and application activities.
This setting enables repeatable collection and links traffic observations to known execution conditions.
However, the resulting traffic does not capture the full diversity of operational networks, where devices, software versions, user behavior, background workloads, and network conditions vary.
The benchmark therefore evaluates classification within the covered collection scenarios; its results do not establish performance on unseen deployment environments.

Label observability also depends on the task and the available traffic representation.
Labels are derived from collection records and task-specific evidence, whereas model inputs contain selected network observations.
A label supported by execution records is not necessarily identifiable from every individual Flow, Flow Window, or Trace.
In particular, an attack-attempt label does not imply successful exploitation, and encrypted traffic may expose behavioral patterns without revealing application content.
Similarly, APT Flow classification evaluates benign/malicious discrimination rather than reconstruction of an entire attack chain.
These distinctions define what each task measures and should guide interpretation of its classification results.




\subsection{Data Access}
\label{subsec:data-access}

TrafficInstruct, including raw packet captures (PCAP files) and supervised fine-tuning (SFT) data, is accessible through the project repository at \url{https://github.com/zhengdao-li/TrafficInstruct}.
The repository provides a dataset overview, download instructions, and links to the cloud-hosted files together with the required access codes.
Download links and access codes are maintained in the repository to accommodate changes in the hosting service.

\section{Conclusion}
\label{sec:conclusion}

We presented TrafficInstruct, a dataset and benchmark comprising raw packet captures and 55,000 SFT records across ten network traffic analysis tasks.
Its task-specific observations, structured supervision, and fixed evaluation protocol support instruction tuning and comparison at coarse and fine label granularities.
Experiments with statistical baselines, ET-BERT, and a 27B language model show that task-specific LoRA adaptation improves Macro-F1 over the unadapted base across all ten tasks, including both plaintext and encrypted Web views.
Differences across tasks and label levels highlight the value of evaluating detection and fine-grained attribution together.
TrafficInstruct provides a common resource for developing and evaluating instruction-tuned models for network traffic analysis.



\clearpage
\appendices
\section{Task Cards and Label Maps}
\label{app:task-cards}

The task cards below describe the collection units, parent sources, labels, and known limitations.
Each task uses the primary Macro-F1 and supplementary metrics specified in Section~\ref{subsec:scoring-rules}.
The fixed data partitions follow the collection-session rule in Section~\ref{subsec:fixed-splits}.
Candidate labels and permitted input fields are specified in Appendix~\ref{app:schema-conversion}; unavailable historical parameters are reported as limitations rather than replaced with current defaults.

\subsection{Malware Traffic Detection}
\label{app:task-malware}

The primary target is benign or one of ten malware families from a connection-level observation; the complete CAPE Run and source sample define its parent groups.

\subsubsection{Collection Procedure and Workflow}
\label{app:collection-malware}
The selected runs record CAPE 2.5 at commit \texttt{fb1cb930713a} and use Windows 10 x64 with INetSim 1.3.2.
The retained dual-VM configuration assigns 8 GiB RAM and four virtual CPUs to each VM.
CAPE restores its configured running snapshot before guest analysis and shuts down the VM afterward.
Host-side \texttt{tcpdump} 4.99.4 captures on the sandbox bridge, selecting the guest's traffic while excluding ARP, guest-agent port 8000, and ResultServer port 2042.
Capture starts after routing is established and before the guest analyzer; cleanup stops it after normal or abnormal analysis termination.
The exact Windows edition and build were not retained.

TCP uses a 120-second idle timeout, ordinary UDP 30 seconds, and UDP DNS 10 seconds with the transaction ID included in its grouping key.
A changed initial SYN endpoint/sequence-number signature separates repeated TCP instances; identical SYN retransmissions remain together unless idle timeout applies.
Noise filtering precedes control identification and the ordered endpoint, process-DNS, sandbox-host, and event-time attribution rules.
Provider-family normalization and CAPE agreement describe sample identity, whereas network-attribution methods describe the evidence for a particular connection.
Among 2,500 malicious records, 1,367 use target-process endpoints, 804 sandbox-host endpoints, 125 application-event timing, and 204 DNS-related timing rules.
These are different forms of support, rather than uniformly direct process confirmation.

The 2,500 benign records come from explicit control runs with synthetic DNS, HTTP/HTTPS, SMTP, FTP, and IRC workloads, not natural Internet background.
Their 87 source runs include 86 PowerShell controls and one historical executable control.
The final selection inherits parent source-group splits and deterministically cycles through sorted source groups within class quotas.
Its 410 complete Runs and source-sample groups are disjoint across the fixed partitions.
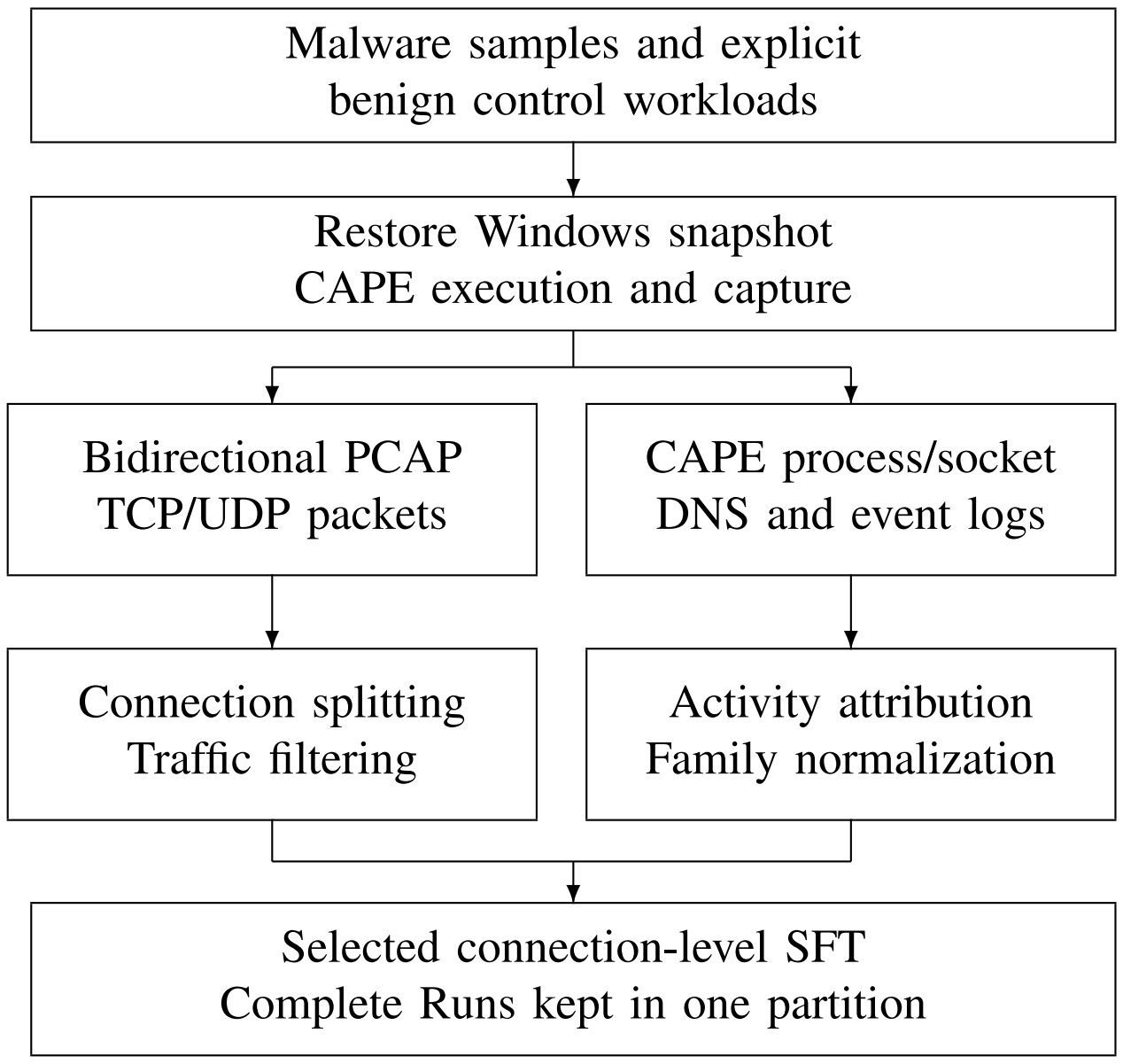
\begin{figure}[!t]
\centering
\begingroup
\setlength{\unitlength}{1pt}
\begin{picture}(250,224)
\put(10,194){\framebox(230,28){\parbox{222pt}{\centering Malware samples and explicit\\benign control workloads}}}
\put(10,154){\framebox(230,28){\parbox{222pt}{\centering Restore Windows snapshot\\CAPE execution and capture}}}
\put(5,102){\framebox(112,36){\parbox{104pt}{\centering Bidirectional PCAP\\TCP/UDP packets}}}
\put(128,102){\framebox(112,36){\parbox{104pt}{\centering CAPE process/socket\\DNS and event logs}}}
\put(5,50){\framebox(112,36){\parbox{104pt}{\centering Connection splitting\\Traffic filtering}}}
\put(128,50){\framebox(112,36){\parbox{104pt}{\centering Activity attribution\\Family normalization}}}
\put(10,0){\framebox(230,32){\parbox{222pt}{\centering Selected connection-level SFT\\Complete Runs kept in one partition}}}
\put(125,194){\vector(0,-1){12}}
\put(125,154){\line(0,-1){8}}
\put(61,146){\line(1,0){123}}
\put(61,146){\vector(0,-1){8}}
\put(184,146){\vector(0,-1){8}}
\put(61,102){\vector(0,-1){16}}
\put(184,102){\vector(0,-1){16}}
\put(61,50){\line(0,-1){9}}
\put(184,50){\line(0,-1){9}}
\put(61,41){\line(1,0){123}}
\put(125,41){\vector(0,-1){9}}
\end{picture}
\endgroup
\caption{Malware collection and connection-level SFT construction. Network attribution rules differ in specificity; the complete CAPE Run and source sample remain the parent sources.}
\label{fig:workflow-malware}
\end{figure}

\subsection{Botnet Detection}
\label{app:task-botnet}

The task classifies each bidirectional Flow as Normal or Botnet at the first level, and as Normal or one of five Botnet families at the second level.
Complete CAPE Runs and source/control groups define its parent sources.

\subsubsection{Collection Procedure and Workflow}
\label{app:collection-botnet}
The collection runs use CAPE 2.5 and the Windows/INetSim snapshot and host-capture mechanism described for malware collection.
Both TCP and UDP use a 120-second idle timeout, without a separate DNS timeout; a different initial SYN signature starts a new TCP instance.
Filtering excludes sandbox management, guest-to-guest communication, and configured Windows discovery services.

Botnet attribution uses process-endpoint matches, capture-level process-DNS matches, and extracted-C2 matches.
Rules select the first satisfied branch in that order.
C2 IP matching does not require port agreement, and DNS rules compare domain sets within a capture rather than reconstructing each query's owning process.
No independent per-Flow manual conflict adjudication is claimed.

Normal traffic comes from explicit clean-control DNS/HTTP scripts.
Final selection uses fixed class quotas and deterministic source/record ordering while retaining parent splits.
Complete Runs and source-sample or control-script groups are kept within a single partition.
\begin{figure}[!t]
\centering
\begingroup
\setlength{\unitlength}{1pt}
\begin{picture}(250,224)
\put(10,194){\framebox(230,28){\parbox{222pt}{\centering Botnet-tagged samples\\and clean DNS/HTTP controls}}}
\put(10,154){\framebox(230,28){\parbox{222pt}{\centering CAPE/INetSim execution\\Capture each complete Run}}}
\put(5,102){\framebox(112,36){\parbox{104pt}{\centering TCP/UDP capture\\Packet observations}}}
\put(128,102){\framebox(112,36){\parbox{104pt}{\centering Process endpoints\\DNS and C2 records}}}
\put(5,50){\framebox(112,36){\parbox{104pt}{\centering Flow splitting\\Noise filtering}}}
\put(128,50){\framebox(112,36){\parbox{104pt}{\centering Normal / Botnet labeling\\Five-family attribution}}}
\put(10,0){\framebox(230,32){\parbox{222pt}{\centering Selected Flow SFT\\Complete Runs kept in one partition}}}
\put(125,194){\vector(0,-1){12}}
\put(125,154){\line(0,-1){8}}
\put(61,146){\line(1,0){123}}
\put(61,146){\vector(0,-1){8}}
\put(184,146){\vector(0,-1){8}}
\put(61,102){\vector(0,-1){16}}
\put(184,102){\vector(0,-1){16}}
\put(61,50){\line(0,-1){9}}
\put(184,50){\line(0,-1){9}}
\put(61,41){\line(1,0){123}}
\put(125,41){\vector(0,-1){9}}
\end{picture}
\endgroup
\caption{Botnet collection and labeling for Normal/Botnet detection and five-family attribution.}
\label{fig:workflow-botnet}
\end{figure}
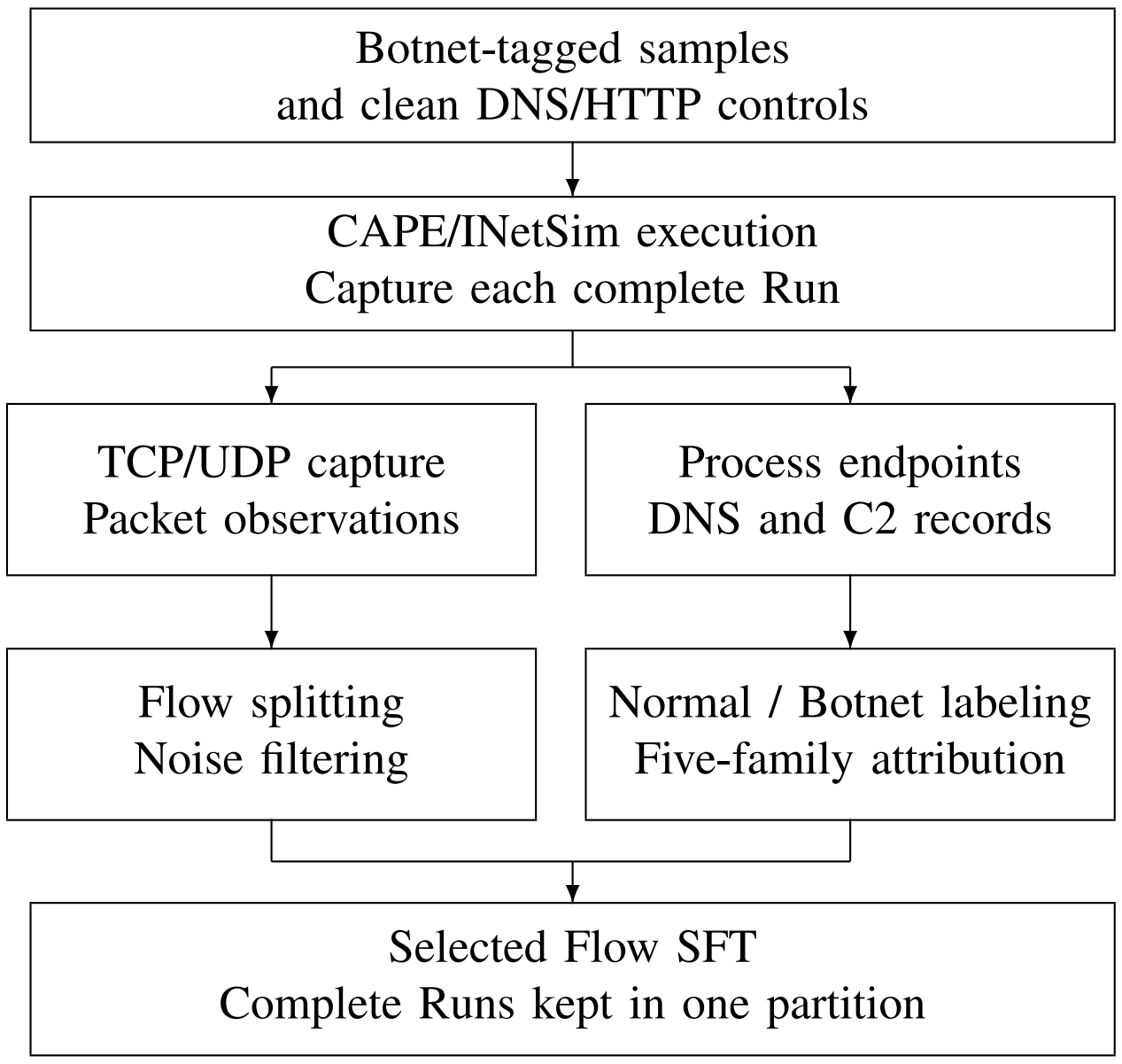

\subsection{Malicious DoH Detection}
\label{app:task-doh}

The task uses Flow feature text for three-class classification of non-DoH, benign DoH, and malicious DoH, and fourteen-class classification that distinguishes twelve malicious categories alongside non-DoH and benign DoH.
Each tunnel execution or browser visit defines its complete source activity.

\subsubsection{Collection Procedure and Workflow}
\label{app:collection-doh}
Three client workers connect to a laboratory DoH server, which forwards DNS to the corresponding iodine, dns2tcp, or dnscat2 server.
The collector selects the worker's experimental interface and assigned TLS port, starting before the client proxy and tunnel establish their connections.
After behavior completion and server-side summary checks, it waits three seconds for remaining traffic before shutting down the tunnel, proxy, and capture.
Worker configurations set a 2,400-second behavior timeout, a 45-second readiness timeout, and at most three attempts.
These limits do not specify the observed duration of every capture.

Browser DoH collection uses fresh Chrome/Firefox profiles, forced resolver settings, and disabled HTTP/3 and ECH.
Page loading has a 45-second timeout, followed by scrolling and a configured dwell period; capture ends after a two-second cooldown.
The non-DoH branch disables browser DoH and excludes a finite list of known DoH endpoints.
Configured whole-session timeout fields of 90/120 seconds were not used by the retained collectors and are not reported as enforced limits.

Malicious candidates require both TLS handshake directions and bidirectional TLS application data, without malformed packets or missing TCP segments; a nonzero capture-kernel drop count also rejects the malicious capture.
Benign checks use resolver SNI and bidirectional application data, along with plaintext-DNS and QUIC checks.
The extractor runs per PCAP with minimum packet count one and no configured time window.
Its feature groups contain six flow-summary values and five groups of 32 packet-length, timing, or burst statistics.
The reported post-check minima of 13/16/14 packets for malicious/benign/non-DoH traffic are observed values, not extractor thresholds.

Exact feature formulas remain to be documented from the feature extractor.
A historical construction script used an inconsistent association between behavior codes M05--M11 and category names.
The reported experiments use the supplied dataset labels; the historical inconsistency is not used to relabel the evaluated records automatically.
\begin{figure}[!t]
\centering
\begingroup
\setlength{\unitlength}{1pt}
\begin{picture}(250,224)
\put(10,194){\framebox(230,28){\parbox{222pt}{\centering Tunnel-tool behaviors or\\configured browser visits}}}
\put(10,154){\framebox(230,28){\parbox{222pt}{\centering Execute activity and capture\\laboratory or browser traffic}}}
\put(5,102){\framebox(112,36){\parbox{104pt}{\centering PCAP / TLS streams\\Packet observations}}}
\put(128,102){\framebox(112,36){\parbox{104pt}{\centering Settings and status\\Resolver checks}}}
\put(5,50){\framebox(112,36){\parbox{104pt}{\centering TLS/data checks\\166 features}}}
\put(128,50){\framebox(112,36){\parbox{104pt}{\centering Task labels\\Coarse and fine categories}}}
\put(10,0){\framebox(230,32){\parbox{222pt}{\centering Feature-text SFT\\Three- and fourteen-class evaluation}}}
\put(125,194){\vector(0,-1){12}}
\put(125,154){\line(0,-1){8}}
\put(61,146){\line(1,0){123}}
\put(61,146){\vector(0,-1){8}}
\put(184,146){\vector(0,-1){8}}
\put(61,102){\vector(0,-1){16}}
\put(184,102){\vector(0,-1){16}}
\put(61,50){\line(0,-1){9}}
\put(184,50){\line(0,-1){9}}
\put(61,41){\line(1,0){123}}
\put(125,41){\vector(0,-1){9}}
\end{picture}
\endgroup
\caption{DoH collection and feature preparation. Fine-category scores use the supplied labels; their behavior-code semantics are discussed separately.}
\label{fig:workflow-doh}
\end{figure}
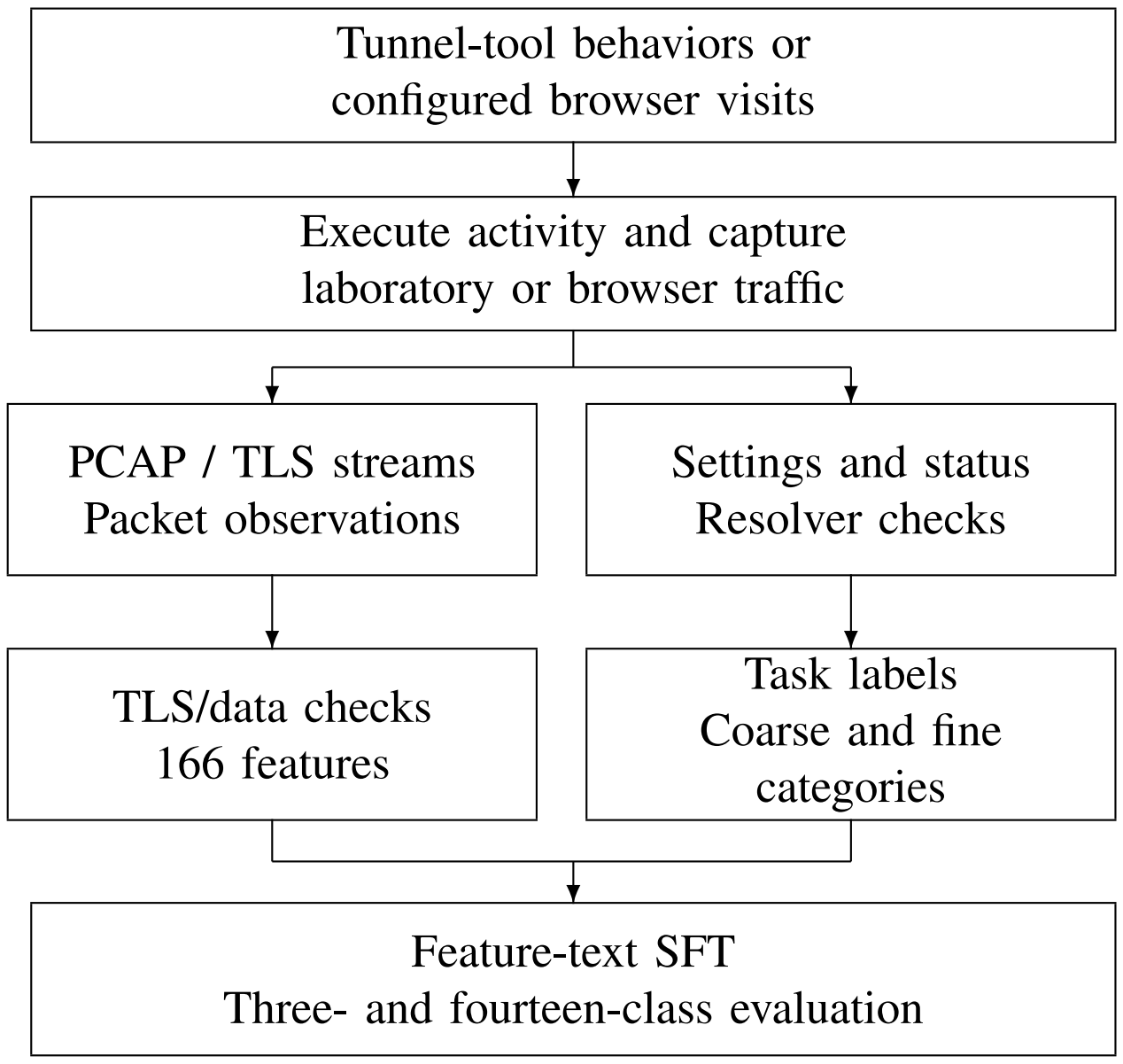

\subsection{Web Attack Detection}
\label{app:task-web}

The task predicts benign traffic or one of eight attempted attack families from a Web Flow.
The dataset provides plaintext HTTP and encrypted TLS views; baseline experiments evaluate the two views separately.
Behavior sessions are nested inside complete capture rounds.

\subsubsection{Collection Procedure and Workflow}
\label{app:collection-web}
\begin{figure}[!t]
\centering
\begingroup
\setlength{\unitlength}{1pt}
\begin{picture}(250,224)
\put(10,194){\framebox(230,28){\parbox{222pt}{\centering OpenClaw attack planning\\and benign activity selection}}}
\put(10,154){\framebox(230,28){\parbox{222pt}{\centering Attack Skills and scripts\\Execute with loopback capture}}}
\put(5,102){\framebox(112,36){\parbox{104pt}{\centering TLS / HTTP captures\\Packet observations}}}
\put(128,102){\framebox(112,36){\parbox{104pt}{\centering Request roles\\Responses and logs}}}
\put(5,50){\framebox(112,36){\parbox{104pt}{\centering TCP Flow parsing\\Event association}}}
\put(128,50){\framebox(112,36){\parbox{104pt}{\centering Attempt labels\\Outcome checks}}}
\put(10,0){\framebox(230,32){\parbox{222pt}{\centering Encrypted and plaintext Flow SFT\\Parent-round and view-pair links}}}
\put(125,194){\vector(0,-1){12}}
\put(125,154){\line(0,-1){8}}
\put(61,146){\line(1,0){123}}
\put(61,146){\vector(0,-1){8}}
\put(184,146){\vector(0,-1){8}}
\put(61,102){\vector(0,-1){16}}
\put(184,102){\vector(0,-1){16}}
\put(61,50){\line(0,-1){9}}
\put(184,50){\line(0,-1){9}}
\put(61,41){\line(1,0){123}}
\put(125,41){\vector(0,-1){9}}
\put(240,68){\line(1,0){7}}
\put(247,68){\line(0,1){140}}
\put(247,208){\vector(-1,0){7}}
\end{picture}
\endgroup
\caption{Web collection with OpenClaw planning and scripted execution. The feedback arrow represents use of experiment results in planning; it does not assert a fresh model call in every round. Attack attempts and local outcomes are separate labels.}
\label{fig:workflow-web}
\end{figure}
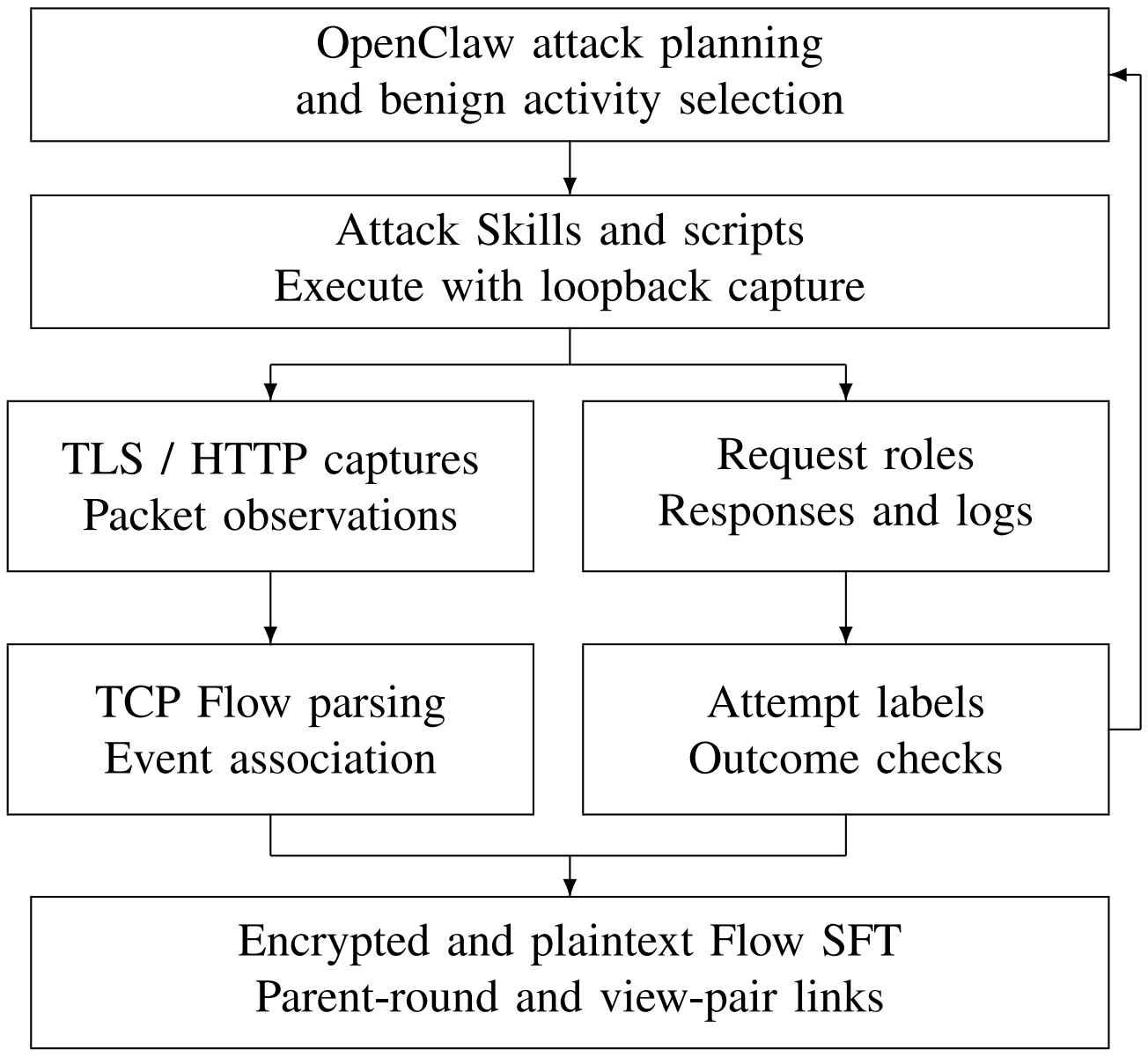

OpenClaw participates in attack planning, while attack Skills and execution scripts carry out the planned interactions with the local OWASP Juice Shop testbed.
Traffic, server-side access/audit logs, and experiment reports provide observations for subsequent planning.
This division of planning and execution does not imply a fresh language-model call in every collection round or a measured improvement from adaptive planning.

A Node HTTPS reverse proxy accepts TLS on port 3443, terminates TLS, and forwards HTTP to a gateway on port 3001; ordinary application requests continue to Juice Shop on port 3000.
The gateway also implements controlled test endpoints.
Two \texttt{tcpdump} processes on the loopback interface capture ports 3443 and 3001.
One collection round contains multiple behavior sessions and shares one pair of capture files.
The selected release represents 229 rounds and 458 view-specific PCAPs.
The parent round and behavior session therefore need separate source identifiers.

Request roles determine labels: attack steps remain malicious even if they fail, whereas normal/context requests are benign.
Any matched attack request makes its Flow malicious; family labels describe the attempted family.
Benign traffic includes homepage visits, searches, static/API requests, and failed logins without an attack payload.
Neither HTTP 200 nor the absence of an error alone establishes attack success.
Controlled outcome checks include reflected XSS markers without browser-script execution, loopback resource retrieval for RFI without remote-code execution, and a whitelisted marker command for command-injection experiments.
These local checks do not demonstrate arbitrary exploitation or deployed IDS/WAF detection.

HTTP association uses method, path, and a three-second matching interval; TLS association uses a nearest request anchor and bounded connection timing within the behavior session.
Connection pools are scoped to client sockets rather than shared globally.
The 5,000 unique view pairs are identified by parent run, behavior session, and complete request-event set, with matching labels and fixed partition assignments.
This is event-based offline association, not independent packet-level causal validation.

Batch checks require valid captures and labels, request-signature repetition no greater than 15\%, attack-outcome success rates within 45--55\%, and success/failure coverage for each family.
The first four source batches passed; the fifth had a 43.75\% success rate and no successful XSS examples, and was included without changing its labels or gate thresholds.
That exception affects outcome quotas rather than converting failed attacks to benign records.
The original plaintext inputs retain experiment paths and markers; these fields must be considered when interpreting plaintext classification results.

\subsection{APT Flow Classification and Multi-Granularity Resources}
\label{app:task-apt}

The primary target is Flow benign/malicious only; 60-second Windows and Run records remain additional linked data, with the complete experiment Run as parent.

\subsubsection{Collection Procedure and Workflow}
\label{app:collection-apt}
The isolated VMware testbed uses Windows nodes with CALDERA Sandcat, Sysmon, and host-side \texttt{dumpcap}; an internal receiving service supports simulated exfiltration.
Nodes have no default route.
Each experiment restores its snapshot, checks service readiness, starts capture, and then launches the planned CALDERA operation when applicable.
Some benign administrative runs do not use CALDERA.
Capture stops normally at the Run's configured duration; no common pre/post buffer or exact software version is inferred where historical records omit it.
Incomplete attempts are quarantined and repeated from a clean snapshot, rather than appended to an admitted Run.

Admission checks require usable captures, consistency with the frozen plan, expected Link counts without skipped Links, and the appropriate success or planned-failure evidence.
The 96 failed or blocked attack Runs exclude infrastructure failures.
These outcomes describe attack execution, not the benign/malicious label of every constituent Flow.
Successful, failed/blocked, and benign experiments contribute 94, 96, and 100 Runs, respectively.
The precise whole-history retry count and per-record manual-review count were not retained.

Protocol and bidirectional endpoints match Sysmon network events to packets first.
Otherwise, UTC stage windows provide fallback labels: the retained single-node script uses delegation minus two seconds through the latest reported completion plus two seconds; the multi-node script uses minus three/plus four seconds.
Overlapping windows select the first matching stage.
Dual-sensor duplicate observations use a 15-ms tolerance; this is not an independently measured clock-offset correction.

Flow keys include protocol, unordered endpoints, and stage, with TCP/UDP idle timeouts of 120/30 seconds and first-packet sender as initiator.
Statistics use all member packets.
The converter supports up to 48 sanitized packet summaries through deterministic head/middle/tail selection, but these lists are empty in the selected Flow records.
The classification baseline uses the available Flow statistics together with retained contextual and identifier fields, without a packet-sequence input.
Nonoverlapping 60-second Windows start at the Run's recorded start, falling back to the first observation only if that field is missing.
Cross-boundary Flows contribute local subflow summaries to each relevant Window; each Window retains at most 64 summaries through deterministic selection.

Fusion can draw on multiple sensor files from the same Run.
Anonymous record/evidence references support tracing; execution states, scenarios, commands, credentials, and markers are not classification features.
The APT baseline evaluates Flow-level benign/malicious classification under the input conditions reported in Section~\ref{subsec:experimental-setup}.
\begin{figure}[!t]
\centering
\begingroup
\setlength{\unitlength}{1pt}
\begin{picture}(250,224)
\put(10,194){\framebox(230,28){\parbox{222pt}{\centering Frozen attack/benign plans\\Restore isolated Windows VMs}}}
\put(10,154){\framebox(230,28){\parbox{222pt}{\centering Start host captures; execute\\CALDERA or benign activities}}}
\put(5,102){\framebox(112,36){\parbox{104pt}{\centering Host-side PCAPs\\Sensor deduplication}}}
\put(128,102){\framebox(112,36){\parbox{104pt}{\centering CALDERA timeline\\Host network events}}}
\put(5,50){\framebox(112,36){\parbox{104pt}{\centering Stage-specific Flows\\Windows and Runs}}}
\put(128,50){\framebox(112,36){\parbox{104pt}{\centering Endpoint/time match\\Run admission}}}
\put(10,0){\framebox(230,32){\parbox{222pt}{\centering Linked multi-granularity SFT\\Initial target: Flow binary classification}}}
\put(125,194){\vector(0,-1){12}}
\put(125,154){\line(0,-1){8}}
\put(61,146){\line(1,0){123}}
\put(61,146){\vector(0,-1){8}}
\put(184,146){\vector(0,-1){8}}
\put(61,102){\vector(0,-1){16}}
\put(184,102){\vector(0,-1){16}}
\put(61,50){\line(0,-1){9}}
\put(184,50){\line(0,-1){9}}
\put(61,41){\line(1,0){123}}
\put(125,41){\vector(0,-1){9}}
\put(128,120){\line(-1,0){6}}
\put(122,120){\line(0,-1){52}}
\put(122,68){\vector(-1,0){5}}
\end{picture}
\endgroup
\caption{APT collection, evidence association, and multi-granularity construction. Windows and Runs remain additional data; the initial benchmark evaluates only Flow benign/malicious classification.}
\label{fig:workflow-apt}
\end{figure}
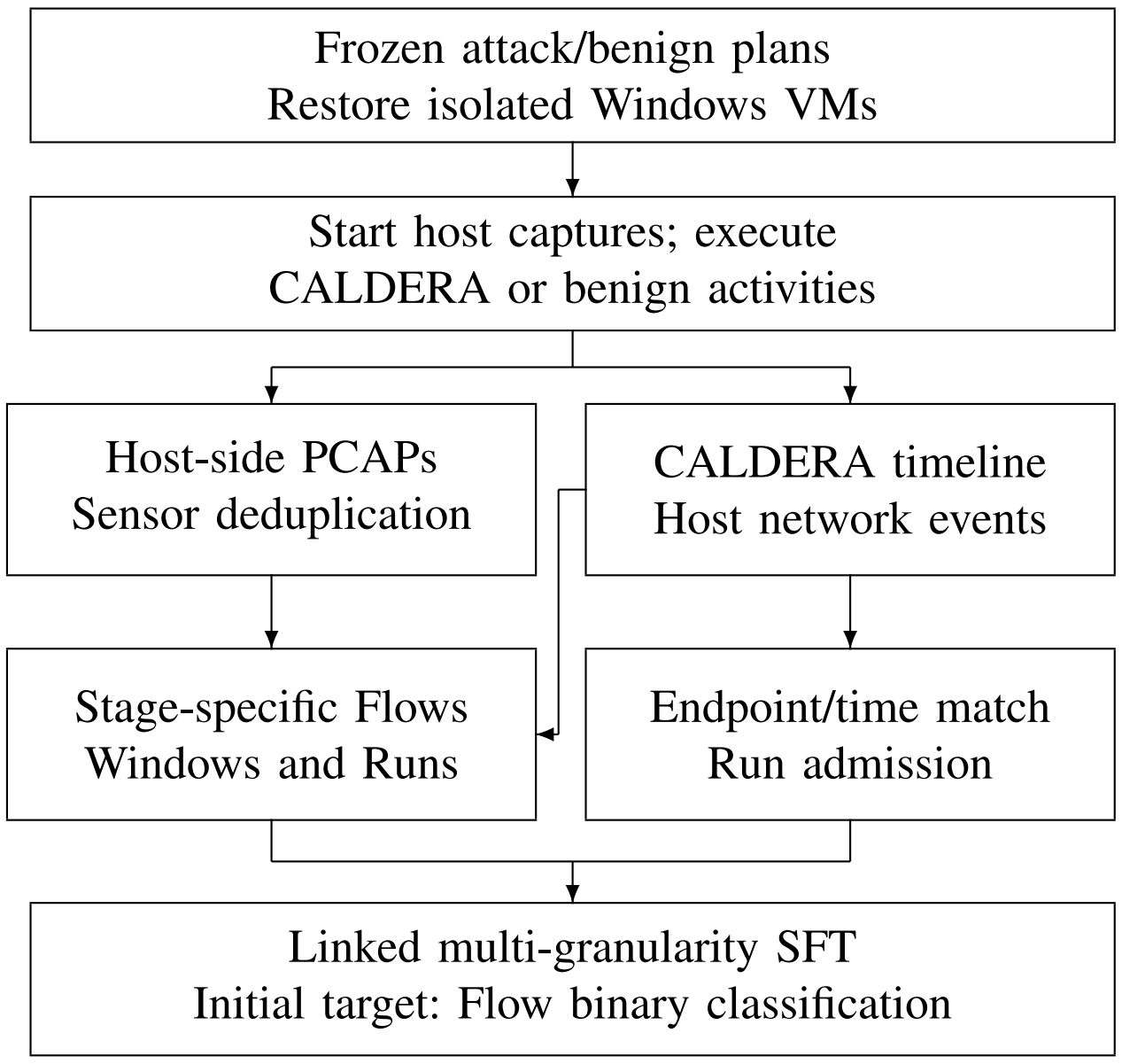

\subsection{Encrypted VPN Detection}
\label{app:task-vpn}

The primary target is one of eighteen communication/activity categories from an encrypted Flow; scenario labels do not assert maliciousness.

\subsubsection{Collection Procedure and Workflow}
\label{app:collection-vpn}

For tunnel-based VPN traffic collection, a Windows host running OpenVPN client version 3.9.0 (build 5008) was connected to a server running the corresponding OpenVPN service.
The client was configured for full-tunnel routing, with UDP as the outer transport for encrypted traffic between the client and server.
For proxy-based collection, a Clash-based client used a virtual TUN interface and global routing with a paid provider's configuration.
The underlying proxy protocol and its cryptographic settings have not been independently verified.
Wireshark 4.6.2 and its capture tools recorded outgoing and incoming network traffic on the capture host, and the packets were saved as PCAP files.

VPN download traffic was generated by downloading files hosted by speed-test websites.
For OpenVPN browsing, a Playwright/Chromium controller automated page visits and scrolling, and page-start and page-end events controlled per-page capture.
Two retained browsing logs associated with released sources each record 200 page visits, with elapsed times of 2,715 and 2,901 seconds.
The retained OpenVPN video collector plays two HLS streams concurrently and is configured for 420-second capture periods with PCAP rotation every 60 seconds.
The retained messaging collector automates message entry after the chat window is focused and uses 1,800-second capture periods with 120-second rotation.
These timing parameters describe the available script configurations; the historical source directories do not establish their use in every collection run.
The released categories also include email and VoIP scenarios.
Available VoIP tools include user-operated call capture and synthetic UDP traffic generation, but their correspondence to individual released captures is incomplete.

Capture interfaces varied by configuration.
Retained logs and sampled source-file headers include WLAN and Ethernet interface labels, and some proxy captures have client-named interfaces.
Interface labels alone do not establish a common physical or virtual capture point for all sources.
The retained OpenVPN browsing, video, and messaging capture commands do not specify a capture-time packet filter; capture and subsequent cleaning are therefore separate steps.
The historical VPN mail-capture logs identify the interface but do not preserve the applied packet filter.

Normal samples were collected by executing application activities directly over the ordinary network, with VPN tunneling and Tor routing disabled.
These activities covered Web browsing, file downloading, email, messaging, video streaming, and VoIP.
The retained control tools implement browser navigation and scrolling, concurrent downloads, HLS playback, and SMTP-over-TLS sending and IMAP-over-TLS retrieval.
For Normal email, a separate PCAP is recorded for each send or retrieval operation, using TCP ports 465 or 993 as the capture filter; one retrieval operation can fetch multiple messages.
This mail filter describes the Normal collector, not a verified filter for the outer VPN connection.
Some Normal samples were reused by the VPN and Tor tasks while retaining their assignment to the same training, validation, or test split.
The Normal label denotes the ordinary-network collection condition; it does not constitute an independent assessment of maliciousness.

The retained Normal browser collector uses process connection records to construct and refresh capture filters, followed by filtering on recorded endpoints and ports.
Capture-time filters exclude broadcast, multicast, and port-53 traffic, with traffic to or from 127.0.0.1 also excluded during warm-up.
UDP discovery ports (5353, 1900, and 5355) are omitted from port-based inclusion lists; other endpoint-matching conditions can still retain such packets.
An optional pre-capture utility can close selected background applications and inspect an idle DNS capture, but its execution is not documented for every collection.
These mechanisms reduce unrelated traffic without establishing that every retained packet belongs to the intended activity.
Available scripts report activity errors and terminate capture at configured stopping conditions; a uniform retry policy and complete attempt-level success or rejection counts have not been established from the retained records.

For both VPN and Tor, we process up to the first 100,000 frames of each selected source PCAP and use TShark TCP/UDP stream identifiers to aggregate packets into bidirectional Flows.
Flows containing fewer than five packets are discarded.
Packets are ordered by timestamp, and at most the first 64 packets of each Flow are retained in the model-visible packet sequence.
Flow-level statistics are computed from all packets of the Flow within the processed portion of the PCAP, before sequence truncation.
Application payload bytes are omitted from the model input.
The partitioning rule is summarized in Appendix~\ref{app:vpn-tor-source-checks}.

\begin{figure}[!t]
\centering
\begingroup
\setlength{\unitlength}{1pt}
\begin{picture}(250,224)
\put(10,194){\framebox(230,28){\parbox{222pt}{\centering OpenVPN/proxy activities\\and ordinary-network controls}}}
\put(10,154){\framebox(230,28){\parbox{222pt}{\centering Execute browsing, transfer,\\streaming, and other activities}}}
\put(5,102){\framebox(112,36){\parbox{104pt}{\centering Capture-host PCAPs\\Packet observations}}}
\put(128,102){\framebox(112,36){\parbox{104pt}{\centering Routing settings\\Activity records}}}
\put(5,50){\framebox(112,36){\parbox{104pt}{\centering Flow aggregation\\Five-packet minimum}}}
\put(128,50){\framebox(112,36){\parbox{104pt}{\centering VPN/Normal labels\\Activity category}}}
\put(10,0){\framebox(230,32){\parbox{222pt}{\centering Encrypted Flow SFT\\Complete activities kept in one partition}}}
\put(125,194){\vector(0,-1){12}}
\put(125,154){\line(0,-1){8}}
\put(61,146){\line(1,0){123}}
\put(61,146){\vector(0,-1){8}}
\put(184,146){\vector(0,-1){8}}
\put(61,102){\vector(0,-1){16}}
\put(184,102){\vector(0,-1){16}}
\put(61,50){\line(0,-1){9}}
\put(184,50){\line(0,-1){9}}
\put(61,41){\line(1,0){123}}
\put(125,41){\vector(0,-1){9}}
\end{picture}
\endgroup
\caption{VPN/proxy and ordinary-network activity collection. All samples from one complete collection activity remain in the same partition.}
\label{fig:workflow-vpn}
\end{figure}
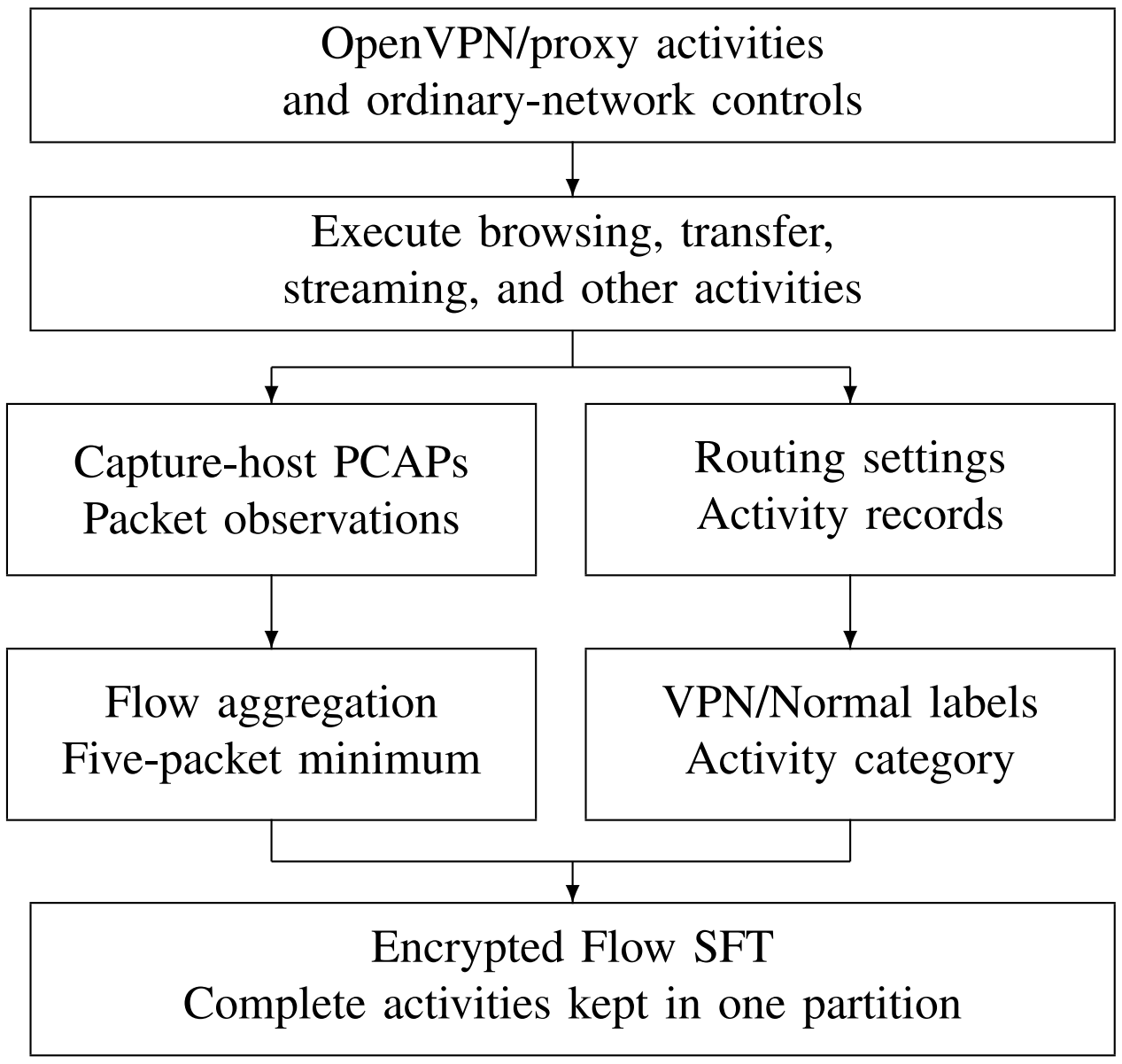

\subsection{Tor Behavior Detection}
\label{app:task-tor}

The primary target is one of twelve communication/activity categories from an encrypted Flow; Normal is an ordinary-network control, not a maliciousness judgment.

\subsubsection{Collection Procedure and Workflow}
\label{app:collection-tor}

Tor traffic was collected under two access settings: access to ordinary Internet services through Tor and access to onion services.
All Tor clients used in the collection were configured to access the Tor network through Snowflake bridges.
Before each new collection session, the preceding Tor circuit was fully closed, and Tor Browser was used to establish a new Tor connection and circuit.
Scripts drove Tor Browser to simulate human browsing behavior; video collection involved opening video webpages and playing their videos.
Other browser activities included file downloading and visiting \texttt{.onion} pages.

Additional scenarios covered anonymous communication and file transfer between two hosts.
For file transfer, one host shared files through an OnionShare service, while the other opened the generated onion-service address in Tor Browser and downloaded the shared files.
Wireshark 4.6.2 ran on one participating Windows host and recorded outgoing and incoming packets at its network interface.
The retained records do not establish whether the file-transfer capture host provided or downloaded the files.
The resulting packet captures were saved as PCAP files for subsequent processing.

For anonymous chat, one host created a chat room using OnionShare, and the other accessed the room through Tor Browser.
The preserved chat-capture directory is described as server-side in its collection notes.
A collection session began with the creation of a new chat room and ended when the corresponding chat communication finished.
After each chat session, Tor Browser and OnionShare were closed and restarted to establish a new Tor connection before a new chat session was initiated.
Splitting a capture into temporal PCAP segments does not define additional collection sessions: the data partitioning protocol requires all derived Flows and samples to inherit their complete parent session's assignment, as specified in Section~\ref{subsec:quality-audits}.

The retained chat and transfer capture scripts select an interface from the host's network-interface enumeration and exclude packets to or from 127.0.0.1, as well as broadcast and multicast traffic, at capture time.
Subsequent filtering uses connection metadata associated with Tor processes.
The chat filtering path and an auxiliary transfer filter also retain packets identified as DTLS or STUN, independently of the endpoint-matching conditions, to include Snowflake communication.
The available filters differ, and their per-capture execution history is incomplete; they do not establish strict process attribution for every retained packet or complete removal of background traffic.
Application-category labels refer to the controlled collection scenario rather than independently verified payload semantics of each Flow.
For example, a Flow captured during chat or file transfer may support Snowflake connection establishment or maintenance without itself carrying chat messages or file contents.

Collection-time Windows, Tor Browser, and OnionShare versions, and per-run timing and retry settings are not completely documented in the retained records.
Complete collection activities define the partitioning units, as described in Appendix~\ref{app:vpn-tor-source-checks}.

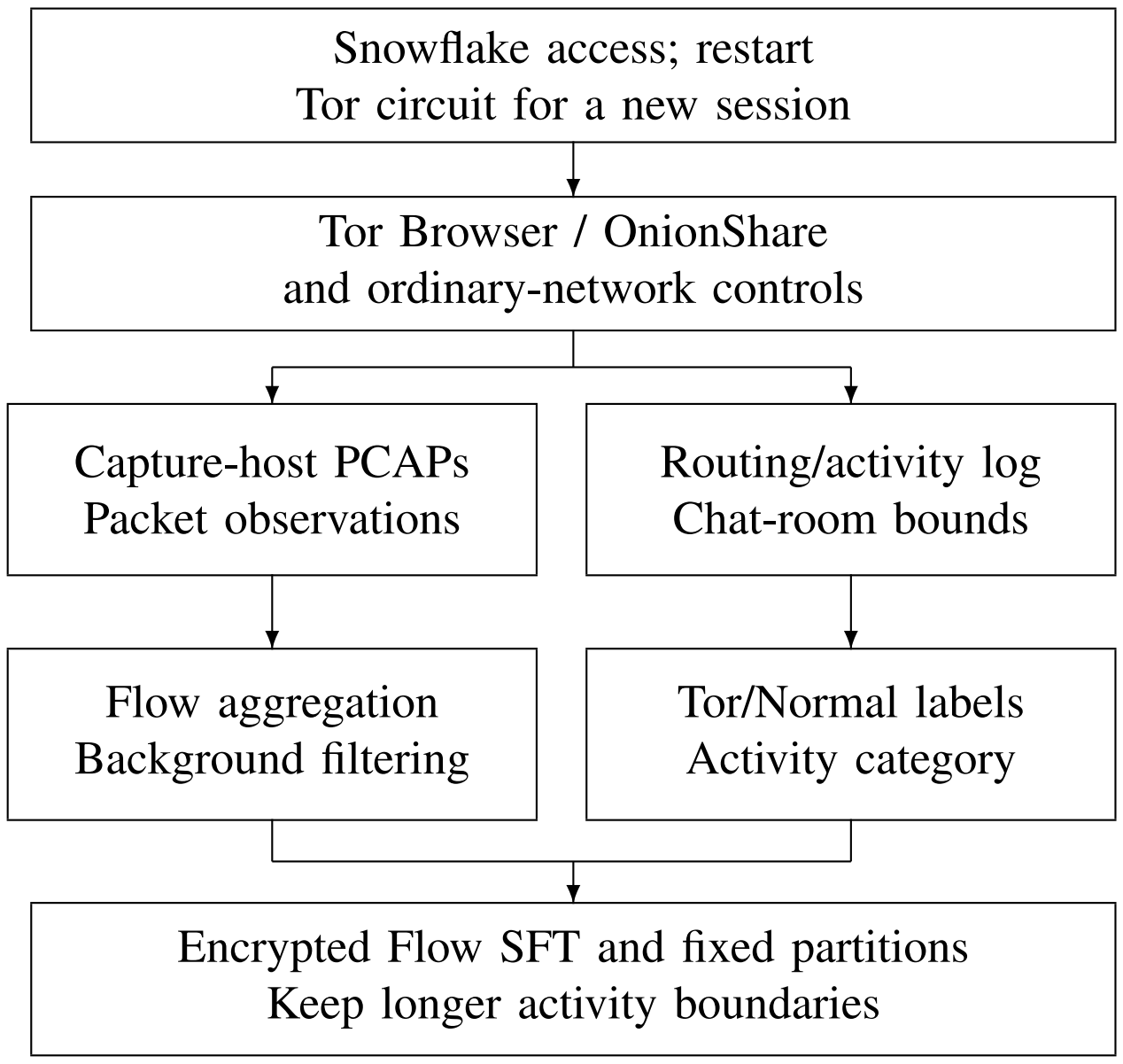
\begin{figure}[!t]
\centering
\begingroup
\setlength{\unitlength}{1pt}
\begin{picture}(250,224)
\put(10,194){\framebox(230,28){\parbox{222pt}{\centering Snowflake access; restart\\Tor circuit for a new session}}}
\put(10,154){\framebox(230,28){\parbox{222pt}{\centering Tor Browser / OnionShare\\and ordinary-network controls}}}
\put(5,102){\framebox(112,36){\parbox{104pt}{\centering Capture-host PCAPs\\Packet observations}}}
\put(128,102){\framebox(112,36){\parbox{104pt}{\centering Routing/activity log\\Chat-room bounds}}}
\put(5,50){\framebox(112,36){\parbox{104pt}{\centering Flow aggregation\\Background filtering}}}
\put(128,50){\framebox(112,36){\parbox{104pt}{\centering Tor/Normal labels\\Activity category}}}
\put(10,0){\framebox(230,32){\parbox{222pt}{\centering Encrypted Flow SFT and fixed partitions\\Keep longer activity boundaries}}}
\put(125,194){\vector(0,-1){12}}
\put(125,154){\line(0,-1){8}}
\put(61,146){\line(1,0){123}}
\put(61,146){\vector(0,-1){8}}
\put(184,146){\vector(0,-1){8}}
\put(61,102){\vector(0,-1){16}}
\put(184,102){\vector(0,-1){16}}
\put(61,50){\line(0,-1){9}}
\put(184,50){\line(0,-1){9}}
\put(61,41){\line(1,0){123}}
\put(125,41){\vector(0,-1){9}}
\end{picture}
\endgroup
\caption{Tor/OnionShare activity collection and encrypted Flow construction. Chat-room boundaries and other recorded source groups are retained without equating every temporal capture segment to a complete Session.}
\label{fig:workflow-tor}
\end{figure}

\subsection{Encrypted Application and Service Classification}
\label{app:task-app}

The primary target is one of 27 application/site/service labels from a Flow Window; a capture fragment is not automatically a complete business session.

\subsubsection{Collection Procedure and Workflow}
\label{app:collection-app}
The 27 labels cover website/resource browsing, video playback and seeking, logged-in social messaging, SSH/SFTP operations, Tor-client requests, and receipt of meeting screen sharing.
They mix applications, sites, and protocol services.
A website label such as a mail or payment service does not imply that mail was sent or a payment performed.
The selected captures cover 47 label/behavior combinations; clients may remain running across capture fragments.

Isolation uses dedicated HTTP/SOCKS or loopback entry points, or observed sockets owned by the target process and its descendants.
The generic process collector captures TCP and polls socket ownership every 0.2 seconds; saved meeting filters additionally identify UDP local ports and bidirectional TCP endpoints.
Proxy establishment is removed through a recognized successful CONNECT response or SOCKS ready response; recognized but incomplete handshakes are rejected.
The retained meeting client version is 3.26.10.401.
Other client and proxy versions are only reported where historical installation or execution records support them; a coded OS-version constant is not treated as a measured environment version.

Capture validity, label/source availability, and behavior-specific quality rules precede conversion.
Continuous activities require at least 80\% of their expected capture duration, while event activities and idle observation windows use separate policies.
Size outliers use within-application/behavior quartiles and a three-IQR rule, with the lower bound clipped at zero.
The final 38 records carrying a manual-acceptance status come from 31 captures; they are not 38 independently reviewed samples.
Nineteen captures retain original reasons, and twelve received a later technical source-filter review rather than reconstructed historical reasons or proof of remote delivery.

TCP/UDP idle timeouts are 120/30 seconds.
Unlike the sandbox tasks, any initial SYN encountered in an active Flow starts a new instance, including an identical SYN retransmission; RST ends a Flow after inclusion, whereas FIN alone does not.
Before adding a packet, a window closes if it already has 128 packets or the next packet is more than ten seconds from its first packet.
Tail windows are emitted and then checked for at least three packets and both directions; all window statistics use its own packets.

Full-capture structure hashes use connection order, direction, IP version, transport, IP and payload lengths, and TCP flags, omitting addresses, ports, payload content, and timing.
Matching capture structures are kept together.
All captures and derived Flow Windows from the same complete business activity are assigned to a single partition.
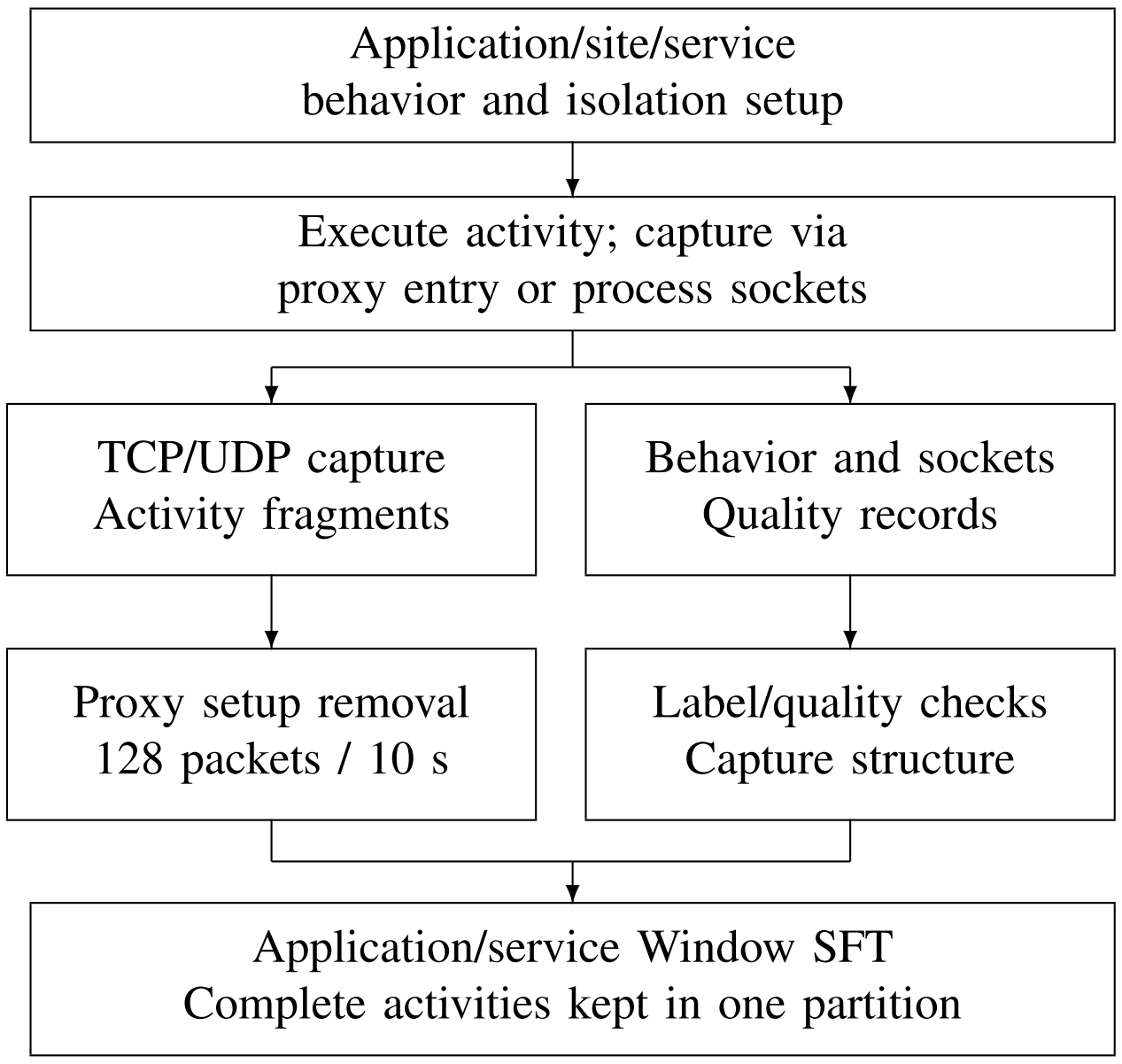
\begin{figure}[!t]
\centering
\begingroup
\setlength{\unitlength}{1pt}
\begin{picture}(250,224)
\put(10,194){\framebox(230,28){\parbox{222pt}{\centering Application/site/service\\behavior and isolation setup}}}
\put(10,154){\framebox(230,28){\parbox{222pt}{\centering Execute activity; capture via\\proxy entry or process sockets}}}
\put(5,102){\framebox(112,36){\parbox{104pt}{\centering TCP/UDP capture\\Activity fragments}}}
\put(128,102){\framebox(112,36){\parbox{104pt}{\centering Behavior and sockets\\Quality records}}}
\put(5,50){\framebox(112,36){\parbox{104pt}{\centering Proxy setup removal\\128 packets / 10 s}}}
\put(128,50){\framebox(112,36){\parbox{104pt}{\centering Label/quality checks\\Capture structure}}}
\put(10,0){\framebox(230,32){\parbox{222pt}{\centering Application/service Window SFT\\Complete activities kept in one partition}}}
\put(125,194){\vector(0,-1){12}}
\put(125,154){\line(0,-1){8}}
\put(61,146){\line(1,0){123}}
\put(61,146){\vector(0,-1){8}}
\put(184,146){\vector(0,-1){8}}
\put(61,102){\vector(0,-1){16}}
\put(184,102){\vector(0,-1){16}}
\put(61,50){\line(0,-1){9}}
\put(184,50){\line(0,-1){9}}
\put(61,41){\line(1,0){123}}
\put(125,41){\vector(0,-1){9}}
\end{picture}
\endgroup
\caption{Application/service collection and Flow Window preparation. Capture fragments from the same complete business activity remain in one partition.}
\label{fig:workflow-app}
\end{figure}

\subsection{Website Fingerprinting}
\label{app:task-wf}

The primary target is an anonymous monitored site or unknown from a complete visit Trace; each independent browser visit is its complete collection Session.

\subsubsection{Collection Procedure and Workflow}
\label{app:collection-wf}
Playwright 1.54.0 drives Chromium revision 1181 (139.0.7258.5), with \texttt{dumpcap} 4.2.2 recorded in the retained environment evidence.
A single worker uses a dedicated Linux network namespace and client interface, capturing TCP/UDP 443 for 30 seconds with a fresh browser context per visit.
Navigation has a 20-second timeout; the configured retry limit permits three total attempts, including the first.
Historical environment evidence does not independently bind every visit to an identical host build.

Acceptance requires error-free navigation, HTTP 200--399, a final hostname matching the requested domain or a subdomain, and at least 20 nonwhitespace body characters.
The title and first 50,000 body characters are checked for configured failure/challenge-page markers.
A usable capture also requires at least 50 packets and both directions.
Failed visits and replacements retain available diagnostics; replacing a monitored site generates new Trace IDs instead of reusing the old visit IDs.

The dataset contains 98 visits to each of 50 monitored sites and one visit to each of 100 unknown domains.
Selection takes a fixed prefix within each parent split/label and restores parent-file order, without an additional random seed.
All 5,000 records correspond to independent accepted visits; unknown domains split 50/25/25 without overlap, whereas monitored domains intentionally recur across splits through different visits.

Packet events follow capture-record order rather than a new timestamp sort.
Time differences are rounded to milliseconds and clipped at zero; signed lengths are IP-packet lengths relative to the configured client direction.
Bursts group consecutive same-direction events without an additional time-gap threshold.
Inputs retain the first 512 events and 128 bursts, while summaries use the full filtered visit and its observed duration rather than a constant 30,000 milliseconds.
\begin{figure}[!t]
\centering
\begingroup
\setlength{\unitlength}{1pt}
\begin{picture}(250,224)
\put(10,194){\framebox(230,28){\parbox{222pt}{\centering Visit schedule and\\monitored/unknown targets}}}
\put(10,154){\framebox(230,28){\parbox{222pt}{\centering Fresh Chromium context\\30-second HTTPS capture}}}
\put(5,102){\framebox(112,36){\parbox{104pt}{\centering Visit PCAP\\TCP/UDP 443 packets}}}
\put(128,102){\framebox(112,36){\parbox{104pt}{\centering Page/domain checks\\Navigation status}}}
\put(5,50){\framebox(112,36){\parbox{104pt}{\centering Events and bursts\\Full-Trace statistics}}}
\put(128,50){\framebox(112,36){\parbox{104pt}{\centering Visit acceptance\\Anonymous site label}}}
\put(10,0){\framebox(230,32){\parbox{222pt}{\centering Visit-Trace SFT with fixed partitions\\Unknown domains kept separate}}}
\put(125,194){\vector(0,-1){12}}
\put(125,154){\line(0,-1){8}}
\put(61,146){\line(1,0){123}}
\put(61,146){\vector(0,-1){8}}
\put(184,146){\vector(0,-1){8}}
\put(61,102){\vector(0,-1){16}}
\put(184,102){\vector(0,-1){16}}
\put(61,50){\line(0,-1){9}}
\put(184,50){\line(0,-1){9}}
\put(61,41){\line(1,0){123}}
\put(125,41){\vector(0,-1){9}}
\end{picture}
\endgroup
\caption{Website visits and Trace preparation. Each accepted fresh-context visit is a complete collection Session; monitored sites recur through separate visits, while unknown domains remain split-disjoint.}
\label{fig:workflow-wf}
\end{figure}
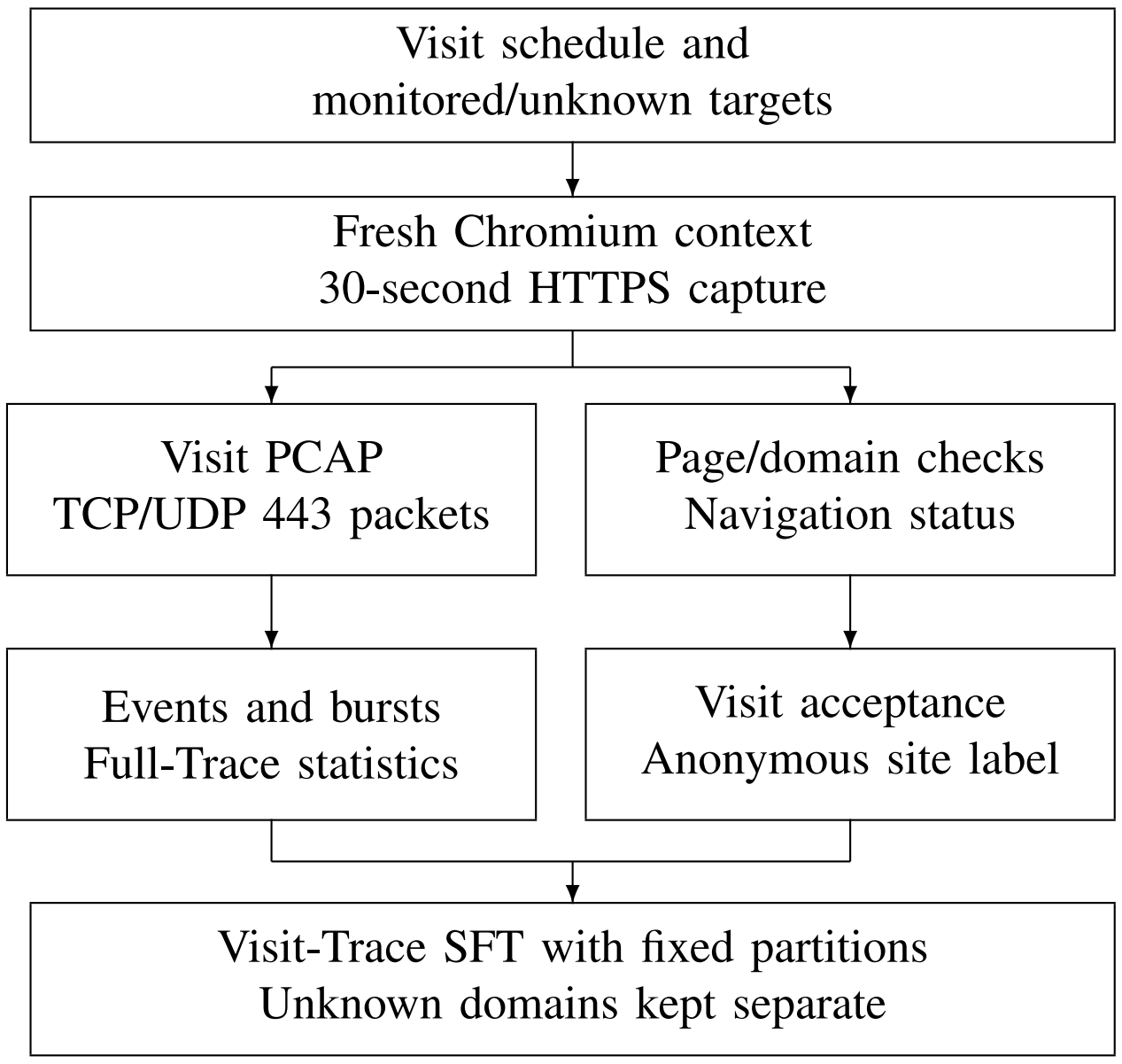

\subsection{App Classification under Concept Drift}
\label{app:task-drift}

The primary target is one of twelve app identities from an attributed Flow; A/B and phase are collection metadata, with each complete registered activity defining its source Session.

\subsubsection{Collection Procedure and Workflow}
\label{app:collection-drift}

Each app has two registered behaviors, A and B, performed separately in individual collection Sessions.
Both behaviors retain the same app classification label.
The behavior designation and collection stage are experiment metadata, not additional target classes.
The registered pairs are:
\begin{itemize}
\item Taobao: A scrolls the homepage once without logging in; B scrolls it twice without logging in.
\item Bilibili: A browses the homepage without logging in, including one fixed scroll; B briefly plays a public video.
\item Weibo: A browses the public homepage feed without logging in, including one fixed scroll; B opens and browses the public video channel.
\item Douyin: A scrolls the recommendation feed once without logging in; B scrolls it twice without logging in.
\item Pinduoduo: A scrolls the homepage once using a user-provided login state; B scrolls it twice using the same login state.
\item Xiaohongshu: A scrolls the homepage/discovery feed once without logging in; B scrolls it twice without logging in.
\item Tencent Video: A plays one public video from the homepage; B plays two public videos consecutively.
\item iQIYI: A plays one public video from the homepage; B plays two public videos consecutively.
\item Amap: A drags the map once at a fixed virtual location; B drags it twice under the same location setting.
\item QQ Music: A plays one public song; B plays two different public songs consecutively.
\item NetEase Cloud Music: A searches without logging in and plays one public song; B searches without logging in and plays two different public songs consecutively.
\item YouTube: A plays one allowlisted public video; B plays two different allowlisted public videos consecutively.
\end{itemize}

The common collection plan specifies 48 Sessions for every app, allocating eight Sessions per app in each of six stages.
P0 executes A only; P1, P2, and P3 allocate respectively six, four, and two Sessions to A and the remaining Sessions to B; P4 executes B only; P5 executes A again in newly collected Sessions.
These are planned behavior counts, not a report of independently verified successful Session totals.
Changing these allocations specifies which activities are executed at each stage; it does not by itself measure how far their traffic distributions differ.

The controller checks completion of the scheduled A/B activities, successful execution of each behavior, and data quality.
Each Session retains its raw capture, app-operation events, and UID/socket records.
Checks cover capture integrity, app-operation records, traffic-to-app matching, bidirectional Flow quality, and the generated JSONL format.
System-shared traffic, traffic with uncertain app ownership, and low-quality Flows are excluded; failed attempts are kept as audit information rather than SFT samples.
The controller arranges additional attempts for missing or unsuccessful Sessions and unsuitable captures.
No concept-drift detection algorithm or traffic-feature threshold is used to establish a drift target or trigger these retries.
\begin{figure}[!t]
\centering
\begingroup
\setlength{\unitlength}{1pt}
\begin{picture}(250,224)
\put(10,194){\framebox(230,28){\parbox{222pt}{\centering Every app: 48 Sessions\\P0--P5, eight per stage}}}
\put(10,154){\framebox(230,28){\parbox{222pt}{\centering Execute registered A/B action\\Capture and observe UID/sockets}}}
\put(5,102){\framebox(112,36){\parbox{104pt}{\centering Device-side PCAP\\TCP/UDP packets}}}
\put(128,102){\framebox(112,36){\parbox{104pt}{\centering UI/media checks\\UID/socket records}}}
\put(5,50){\framebox(112,36){\parbox{104pt}{\centering Unique app ownership\\Flow filtering}}}
\put(128,50){\framebox(112,36){\parbox{104pt}{\centering Execution and quality\\Schedule completion}}}
\put(10,0){\framebox(230,32){\parbox{222pt}{\centering App-label SFT and fixed partitions\\No feature-based drift detector}}}
\put(125,194){\vector(0,-1){12}}
\put(125,154){\line(0,-1){8}}
\put(61,146){\line(1,0){123}}
\put(61,146){\vector(0,-1){8}}
\put(184,146){\vector(0,-1){8}}
\put(61,102){\vector(0,-1){16}}
\put(184,102){\vector(0,-1){16}}
\put(61,50){\line(0,-1){9}}
\put(184,50){\line(0,-1){9}}
\put(61,41){\line(1,0){123}}
\put(125,41){\vector(0,-1){9}}
\put(240,68){\line(1,0){7}}
\put(247,68){\line(0,1){140}}
\put(247,208){\vector(-1,0){7}}
\put(128,120){\line(-1,0){6}}
\put(122,120){\line(0,-1){52}}
\put(122,68){\vector(-1,0){5}}
\end{picture}
\endgroup
\caption{The common 48-Session-per-app collection plan and quality-based retries. The feedback arrow denotes missing or failed activities and unsuitable data, not a statistical drift detector or adaptation of app/network conditions.}
\label{fig:workflow-drift}
\end{figure}
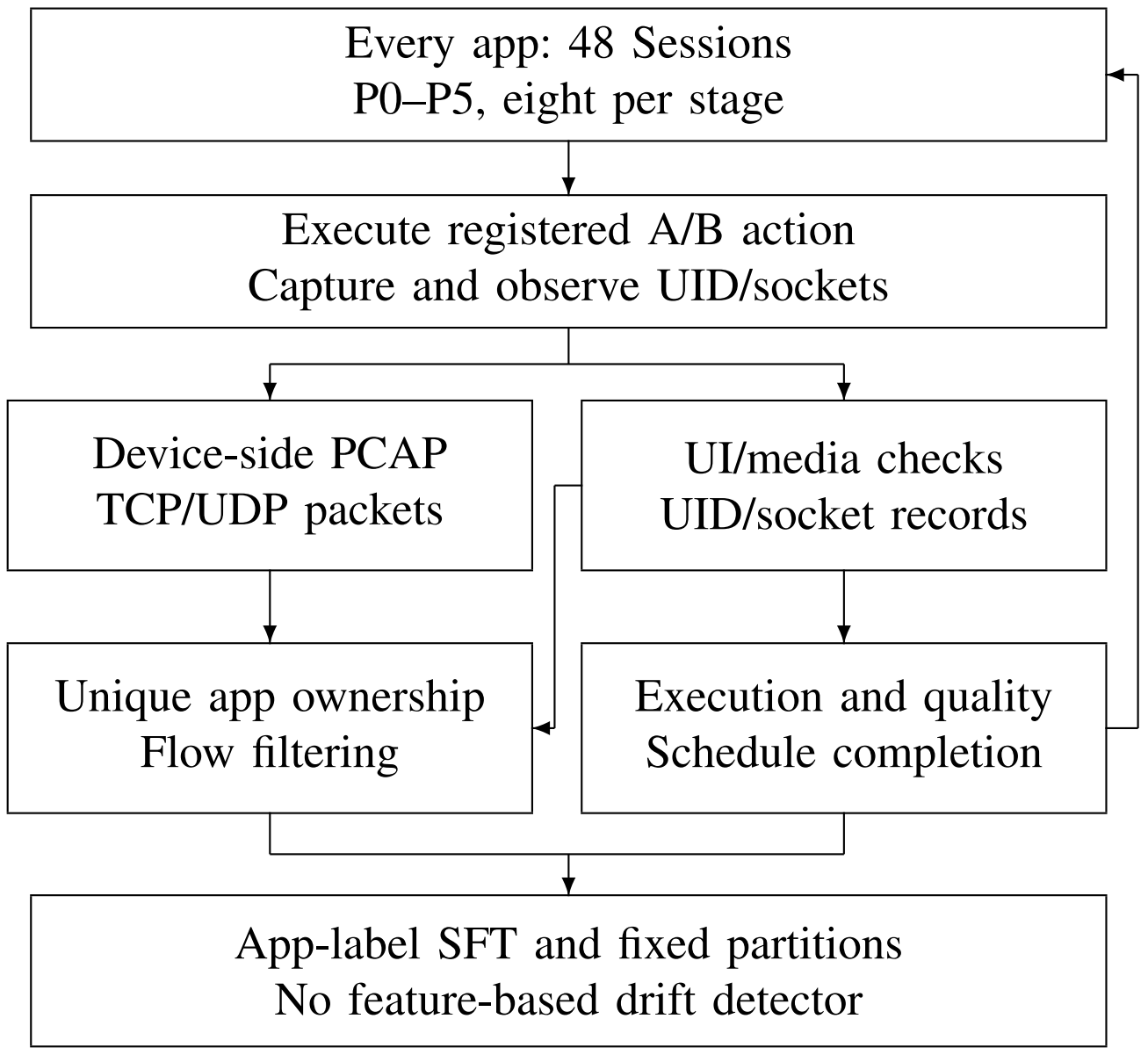

Collection uses Android 13/API 33 arm64 emulators hosted on Apple Silicon macOS.
The device-side capture listens on \texttt{any} for TCP/UDP, writes a temporary PCAP, and transfers it to the host through ADB after capture.
UID/socket observation runs in parallel with capture; fixed UI scripts start after capture initialization.
Capture duration, behavior waiting time, and the registration's nominal duration are different settings and are not presented as one universal time limit.
App data and caches are retained; force-stopping and restarting an app does not clear its state.
Pinduoduo uses the authorized user-provided login state, and YouTube uses the dedicated emulator/network arrangement rather than a claimed common direct route for all apps.

Operation checks use foreground activity, UI state, screenshots, action counts, or media-session playback state as appropriate.
They do not prove full media download or completion of every remote interaction.
UID ownership matches remote endpoints and time windows, using a one-second matching tolerance by default.
Candidates require unique ownership, at least five packets, and at least one packet in each direction.
The retained converter uses a 120-second idle timeout for both TCP and UDP; up to 256 packet summaries are retained while full-Flow statistics remain available.
Here, \texttt{high} confidence denotes a rule-based attribution category, not a calibrated probability.
Flows with insufficient packets or uncertain ownership are removed individually; an otherwise usable Session need not be retried because some of its Flows were rejected.

The dataset assigns P0--P4 to training/validation and P5 to testing, with all Flows from one complete Session kept together.
The collection plan specifies 48 Sessions per app. Samples are selected after quality filtering, while preserving the stage holdout and complete-session partitioning.
The fixed partitions follow Section~\ref{subsec:quality-audits}.
The model predicts app identity from the permitted network features, while A/B designations, phase assignments, and execution records describe how the data were collected.

\section{Schema and Conversion Details}
\label{app:schema-conversion}

\subsection{Task-Specific Answers and Metric Extraction}
\label{app:label-projection}

The task-specific answer formats and scoring-field extraction rules are summarized below. Experimental settings and results are reported in Section~\ref{sec:baseline-experiments}.
The existing instruction/input/output records already define the training tasks.
Nine tasks retain their instructions and original structured answers by default; loaders adapt them to the selected model and scorers read the relevant fields.
Table~\ref{tab:task-output-schemas} separates these answer formats from the internal class used for Macro-F1.
The scorer extracts a class for evaluation without rewriting the supervised answer.

\begin{table*}[!t]
\caption{Task-specific model answers and main-score extraction. The nine retained formats are already present in the SFT data. Only APT uses a scoped binary answer for benchmark evaluation. Auxiliary/schema checks are separate from primary class correctness.}
\label{tab:task-output-schemas}
\centering
\small
\setlength{\tabcolsep}{4pt}
\renewcommand{\arraystretch}{1.25}
\begin{tabular*}{0.9\textwidth}{@{\extracolsep{\fill}}>{\centering\arraybackslash}m{0.09\textwidth}>{\raggedright\arraybackslash}m{0.34\textwidth}>{\raggedright\arraybackslash}m{0.41\textwidth}@{}}
\hline
\textbf{Task} & \textbf{Model-answer fields} & \textbf{Main-score extraction} \\
\hline
MTD & \texttt{maliciousness}, \texttt{malware\_family} & Benign, or family when predicted malicious \\
BND & \texttt{traffic\_class}, \texttt{fine\_label}, \texttt{botnet\_family} & \texttt{fine\_label} \\
MDD & \texttt{label}, \texttt{category} & Coarse class for non-malicious cases; behavior category for malicious \\
WAD & \texttt{label}, \texttt{category} & \texttt{category}; map none to benign \\
AAD & Derived \texttt{label} only & Benign/malicious from the scoped answer \\
EVD / TBD & \texttt{label}, \texttt{category} & \texttt{category} \\
EAC & \texttt{label}, \texttt{label\_id} & Textual \texttt{label} \\
WF & \texttt{site\_label}, \texttt{is\_monitored} & \texttt{site\_label} \\
CD & \texttt{label}, \texttt{label\_id} & Package-name \texttt{label} \\
\hline
\end{tabular*}
\end{table*}

For malware, a valid \texttt{benign} value in \texttt{maliciousness} maps to \texttt{benign}.
A valid \texttt{malicious} value additionally requires \texttt{malware\_family} to be ACRStealer, AgentTesla, AmateraStealer, AsyncRAT, IcedID, Lumma, NanoCore, NetWire, RedLine, or Vidar.
Missing, wrongly typed, or unrecognized \texttt{maliciousness} makes the primary prediction invalid, even if a valid family is supplied.
Botnet uses Normal, AsyncRAT, Lumma, SnakeKeylogger, ZLoader, and njRAT as its fine-label vocabulary.
Normal maps to the coarse class Normal, and each family maps to Botnet.
The generated coarse-class and family fields are checked separately for consistency.
For Web, the primary prediction reads only \texttt{category}: \texttt{none} maps to \texttt{benign}, while the other valid tokens identify SQL injection, XSS, LFI, RFI, command injection, path traversal, Web scanning, or brute force.
The separate \texttt{label} field is checked for auxiliary correctness and consistency; its errors do not invalidate a valid primary category, and it cannot repair an invalid or missing category.
The original APT instruction requests stages, techniques, evidence, and uncertainty, which exceeds the initial Flow binary target.
Its classification baseline therefore uses a dedicated instruction asking for benign/malicious classification from the supplied Flow-level input and a JSON answer containing only \texttt{label}, taken from \texttt{output.assessment.label}.
No Window/Run answer or attack-success class is projected into this task, and the original rich APT records are retained.

For DoH, the stored coarse label distinguishes non-DoH, benign DoH, and malicious DoH.
Fine-label evaluation uses the supplied twelve malicious category names together with non-DoH and benign DoH.
The same label vocabulary is used throughout training, validation, and testing.
The \texttt{label}/\texttt{category} output remains intact; a limited instruction clarification aligns its field names and coarse tokens with the stored answers.
The coarse aliases are explicit and do not authorize reinterpretation of arbitrary generated text.

VPN and Tor use the exact released category strings, including their existing case, rather than inferring labels by unrestricted case normalization.
Their fixed mappings recover the coarse communication class.
Application/service labels use the 27 strings and mobile-app labels use the twelve package names in the supplied data.
Their \texttt{label\_id} fields remain part of the training answers, but ID correctness and consistency do not alter the primary textual-label score.
Website labels are w001--w050 and unknown; the original \texttt{is\_monitored} field is retained and checked against the site label.

Any invalid reference combination is a construction issue to resolve and record before evaluation, not a reason to silently drop a test error.
Predictions use task-specific field extraction and the fixed candidate set under Section~\ref{subsec:scoring-rules}.
Malformed JSON or an invalid/missing scoring field is a primary prediction error; auxiliary fields cannot repair the primary prediction.
No reasoning explanation or evidence ID is generated as part of the initial classification target.

\subsection{Permitted Classification Inputs}
\label{app:input-policy}

Unknown fields are not included automatically when a source schema changes.
Schema/view/encoding descriptions may be supplied once at task level.
Reference answers are prepared separately from model inputs. Input fields are task-specific; the APT experiments retain the contextual and identifier fields described below.
Sequence directions and relative times retain their supplied interpretation, rather than being recomputed from removed endpoint identities.

For malware, permitted fields cover transport/application protocol and service port, duration and packet/IP/payload-byte totals, directional statistics, packet-length/interarrival/burst statistics, and the sanitized packet sequence.
Existing DNS counts, HTTP method/status counts, TLS record/version information, and coarse payload-statistic buckets may be retained, without DNS names or payload text.
Botnet additionally retains its supplied sequence count/selection flags and packet/header fields; sample identity and attribution records are excluded.

VPN, Tor, and staged mobile-app inputs retain their supplied protocol, duration/count, direction, length, relative timing, transport-header, and TLS-record fields.
Empty payload fields and their encoding declarations are not reinstated as content.
Plaintext Web experiments use the compressed HTTP representation, including the payload content retained by the input conversion.
Encrypted Web experiments use the TLS traffic representation without plaintext application payloads.
Application/service records retain the sanitized Window statistics and packet/TLS structure without restoring endpoint names or ports removed upstream.
Website records retain ordered event and burst arrays, their encoding/sequence metadata, and complete-visit statistics.
DoH retains its named feature text and values, with exact extractor definitions still a reproducibility gap.

APT inputs retain the Flow-level statistics and the contextual and identifier fields present in the experimental input objects, including Flow identifiers, local-context references, and evidence-candidate identifiers.
The selected packet-summary lists are empty, so no sequence is synthesized from the other granularities.
These inputs should not be interpreted as identifier-free Flow statistics.

Loaders must validate reference labels and field membership without supplying reference outputs to $G_t$.
They distinguish missing from inapplicable values and document conversions and any extra sequence shortening.
Model-specific serialization and sequence-length limits are applied during input preparation; the experimental settings are reported in Section~\ref{subsec:experimental-setup}.

\section{Data Partitioning}
\label{app:source-audits}
\label{app:delivery-source-checks}
\label{app:evaluation-version}
\label{app:vpn-tor-source-checks}

We verified the collection source of every sample and assigned all observations from the same complete collection session to one partition.
Linked Web views, capture fragments from one activity, and APT observations derived from the same Run retain a common partition assignment.
For mobile-app classification, training and validation use P0--P4, while testing uses P5.
VPN and Tor share Normal controls within matching partitions; these shared observations make their task results correlated without placing identical inputs across training, validation, and test.
Collection-session isolation prevents leakage from the same activity but does not establish generalization to unseen devices, networks, or software environments.

\end{document}